\documentclass[12pt]{article}
\usepackage{epsfig}     

\def\be               {\begin{equation}}
\def\ee               {\end{equation}}
\def\bea              {\begin{eqnarray}}
\def\eea              {\end{eqnarray}}

\def\nn               {\nonumber}

\def\ma[#1,#2,#3,#4]  {{\left( \matrix{ #1  & #2 \cr
                                        #3  & #4 \cr } \right)}}

\begin{document}

\title{
{\vspace{-0cm} \normalsize
\hfill \parbox{40mm}{CERN -TH/98-127}\\
\hfill \parbox{40mm}{MPI-PhT/98-34}}\\[25mm]
Ordering monomial factors of polynomials in the product representation}                
\author{
B. Bunk$^1$,  
S. Elser$^1$, 
R. Frezzotti$^2$
and K. Jansen$^3$  
\\
{\footnotesize 
$^1$ Institut f\"ur Physik, Humboldt--Universit\"at, Invalidenstr. 110,
10115 Berlin, Germany 
}
\\
{\footnotesize 
$^2$ Max-Planck-Institut f\"ur Physik, F\"ohringer Ring 6, D-80805
     M\"unchen, Germany
}
\\
{\footnotesize 
$^3$ CERN, 1211 Gen\`eve 23, Switzerland  
}
}
\date{\today}
\maketitle

\begin{abstract}
The numerical construction
of polynomials
in the product representation (as used for instance in variants of the multiboson technique) 
can become problematic
if rounding errors induce an imprecise
or even unstable evaluation of the polynomial. 
We give criteria to quantify the effects of these rounding errors on the
computation of polynomials approximating the function $1/s$. We consider
polynomials both in a real variable $s$ and in a Hermitian matrix.  
By investigating several ordering schemes for the monomials of these
polynomials, we finally demonstrate that there exist orderings of 
the monomials that keep rounding errors at a tolerable level. 

\end{abstract}

\section{Introduction}

In Monte Carlo simulations of fermionic systems in a discretized
space-time, 
the determinant of a matrix $A$, related to the lattice action of the
fermions,  usually has to be taken into account. 
Although the form of the matrix $A$ can be very general and depends on the
problem under consideration, we assume in the following that $A$
defines a {\em local} action of the fermions extending only over a few 
lattice spacings.
A standard way to incorporate
${\rm det}A$ into simulation algorithms is to write it as a Gaussian integral
\begin{equation} \label{start}
{\rm det} A = \int {\cal D}\Phi^\dagger {\cal D}\Phi e^{-\Phi^\dagger A^{-1} \Phi }
\end{equation}
where $\Phi$ is a suitable complex $N$-component vector on which the matrix $A$ acts and 
${\cal D}\Phi^\dagger {\cal D}\Phi$ the corresponding 
(properly normalized) integration measure. 
One is therefore led to the 
inconvenient problem of inverting an $N\otimes N$ matrix, 
which can be very large, 
i.e. $N$ being ${\rm O}(10^6)$.

In \cite{lue94} an alternative approach was introduced: 
the
determinant of $A$ may be approximated by the inverse determinant of a 
polynomial $P_{n}(A)$ of degree $n$ in the matrix $A$ such that
\begin{equation} \label{p}
{\rm det} A \approx \left[ {\rm det} P_{n}(A) \right]^{-1}\; . 
\end{equation}
In eq.~(\ref{p}) and throughout the rest of the paper we assume that $A$ is 
Hermitian, positive definite, 
and that $\|A\|\le 1$, where 
$\|A\|$ is given by the largest eigenvalue $\lambda_{\rm max}(A)$
of $A$.
In the following, we will consider only polynomials of even degree $n$.
The roots $z_k$, $k=1,...,n$, of the polynomial hence come 
in complex-conjugate pairs and the determinant ${\rm det}P_n$ can be factorized
into positive factors,
resulting in 
\begin{equation} \label{pfact}
\left[ {\rm det} P_{n}(A) \right]^{-1} \propto 
\prod_{k=1}^{n/2} \left[ |{\rm det} (A - z_k)|^2 \right]^{-1} \propto
\prod_{k=1}^{n/2}\int {\cal D}\Phi^\dagger_k {\cal D}\Phi_k 
e^{-\sum_k \Phi^\dagger_k (A-z_k)^\dagger (A-z_k) \Phi_k }\; .
\end{equation} 
From
eq.~(\ref{pfact}), we can see that now the action of the bosonic fields $\Phi_k$ is
local and hence the task of inverting the matrix $A$ can be avoided. 
Similar steps lead
to the so-called multiboson technique for simulating fermionic systems. 
This technique has been shown to
give a comparable performance 
$[$2--8$]$ 
to the standard Hybrid Monte Carlo (HMC) \cite{hmc} or Kramers equation \cite{horowitz,kramers} 
simulation algorithms. Examples for applications of the multiboson technique are  
Monte Carlo simulations of lattice QCD \cite{forcrand3,karlreview}, supersymmetry
\cite{susy},
the Schwinger model \cite{elser} and 
the 
Hubbard model \cite{sawicki}. 

In exact versions of the multiboson technique \cite{galli} 
or related approaches \cite{taka,frezzi,PHMC_hitech}  
the use of a product representation 
of the polynomial $P_n(A)$ often turns out to be convenient. 
However, 
the numerical construction 
of a polynomial using the
product representation can
--because of rounding errors-- 
easily lead to a loss of precision or even to numerical instabilities.
This holds true in particular if
computers with 32-bit arithmetic precision
are used. 
Motivated by simulation algorithm studies, which use the product 
representation of a polynomial, we will show in this paper that by suitable 
orderings of the monomial factors, the precision losses can be kept
on a tolerable level. Although we will only demonstrate this for 
a particular example of the matrix $A$, we expect that similar results may also 
hold for more general situations.

The paper is organized as follows: 
in section 2 we will specify the polynomial we have used in this work.
Effects and possible origins of rounding errors when the polynomial
is evaluated in its product representation are discussed. In section 3
we give the ordering schemes for the monomials of the polynomial
in the product representation. We show how the evaluation of the polynomial 
is affected when the different ordering schemes are employed. 
Section 4 is devoted to quantitative {estimates} of the rounding-error effects.
We compare in section 5 some results when using 32-bit and 64-bit arithmetics
and conclude in section 6.

\section{Product representations of polynomials and rounding errors} 

Let us consider the approximation of a function $f(s)$
depending on a real variable
$s>0$
by a polynomial $P_n(s)$ of degree $n$.                       
The motivation 
to initially study
a single degree of freedom, is 
--besides its simplicity-- that we might think of the matrix $A$ as being diagonalized.
Then the problem, eq.~(\ref{p}), reduces to finding
 a polynomial that approximates
each $\lambda^{-1}(A)$ separately, where $\lambda(A)$ is a real eigenvalue of $A$. 
We therefore expect that studying a single degree of freedom can 
provide information also about the
qualitative behaviour of rounding-error effects when the polynomial $P_n(A)$
in the matrix $A$ is computed. 

To be specific, we follow the 
Chebyshev approximation method \cite{foxy,numrec} 
and construct a polynomial       
approximating the function
$f(s)=1/s$
in the range $\epsilon\le s\le 1$, where $\epsilon\ge 0$ is an adjustable parameter.
The choice of the function $1/s$ is, of course, 
motivated by our original problem of evaluating an inverse matrix, eq.~(\ref{start}). 
We take the same polynomial $P_{n,\epsilon}(s)$ as specified in \cite{bunk}. 
For completeness, we recall here its definition. 
If we introduce the scaled variables $u$ and $\theta$  
\begin{equation} \label{uandtheta}
u= (s-\epsilon)/(1-\epsilon)\;, \;\; \cos\theta = 2u-1
\end{equation}
the Chebyshev polynomial $T_r^{*}(u)$ of degree $r$ is given by
(note that $T_r^{*}$ is {\em not} the standard definition of the Chebyshev
polynomial):
\begin{equation} \label{T_r} 
T_r^{*}(u) = \cos(r\theta)\; .
\end{equation} 
For given $n$ and $\epsilon$ the polynomial $P_{n,\epsilon}(s)$ is then defined by
\begin{equation} \label{THEpolynomial}
P_{n,\epsilon}(s) = \left[1+\rho T_{n+1}^{*}(u)\right]/s\; ,
\end{equation} 
where 
the constant $\rho$ has to be taken such that the square bracket vanishes
at $s=0$. 
The polynomial eq.~(\ref{THEpolynomial}) approximates the function $1/s$ uniformly in the
interval $\epsilon \le s \le 1$.
The relative fit error
\begin{equation}
R_{n,\epsilon}(s) = \left[ P_{n,\epsilon}(s)-1/s \right] s
\end{equation}
in this interval decreases exponentially with increasing $n$
\begin{equation} \label{accuracy}
|R_{n,\epsilon}(s)| \le \delta \equiv 2 
\left(\frac{1-\sqrt{\epsilon}}{1+\sqrt{\epsilon}}\right)^{n+1}
\; .
\end{equation}
The accuracy parameter $\delta$ 
provides an upper bound for the absolute value of the relative fit error 
in the given interval. We note in passing that $|\rho|\le\delta$ and that
in all the practical applications studied in this paper the value
of $\rho$ is actually very close to the value of $\delta$.

The roots $z_k$ of the polynomial 
eq.~(\ref{THEpolynomial}) can be computed analytically,    
\begin{equation} \label{roots}
z_k = \frac{1}{2}(1+\epsilon) - \frac{1}{2}(1+\epsilon)\cos\left(\frac{2\pi k}{n+1}\right)
    - i\sqrt{\epsilon}\sin\left(\frac{2\pi k}{n+1}\right)\; .
\end{equation}
We then obtain the desired product representation of the polynomial
\begin{equation} \label{pnepsilon}
 P_{n,\epsilon}(s)=\prod_{k=1}^n [c_k(s-z_k)]\; .  
\end{equation} 
The real normalization factors $c_k$ 
have to satisfy the condition 
\begin{equation} \label{ck}
\prod_{k=1}^n c_k
=
C_{\rm tot} (n)
=
\left( {1+\epsilon \over 2}
\prod_{k=1}^n \left[ {1+\epsilon \over 2}-z_k\right] \right)^{-1}\; .
\end{equation}
When the $c_k$'s are taken to be all identical, they turn out to be $O(1)$. 
We will also use later the partial products
\begin{equation} \label{plepsilon}
 P_{n,\epsilon}^{l}(s)=\prod_{k=1}^l [c_k(s-z_k)]\; .  
\end{equation} 

One may also be interested \cite{PHMC_hitech} in 
a product representation of
the polynomial  $P_{n,\epsilon}(s)=p_{n,\epsilon}(q)$ in terms 
of the real variable $q=\pm \sqrt{s}$:
\be \label{pol_prod_q_start}
p_{n,\epsilon}(q) = \prod_{k=1}^{2n} [\sqrt{c_k}(q-r_k)] \; ,
\ee
where the roots $r_k$ are defined as 
\begin{eqnarray} \label{r_k} 
r_k & = & \sqrt{z_k},\; {\rm Im}(\sqrt{z_k}) > 0\; ,\; k=1,\dots ,n  \nonumber \\ 
r_k & = & r_{2n+1-k}^{*} \; , \;\;\; k=n+1, \dots, 2n\; 
\end{eqnarray}
with the obvious generalization for the constants $\sqrt{c_k}$.
The representation of eq.(\ref{pol_prod_q_start}) is used in the algorithm
described in \cite{frezzi}, where it leads to a most efficient 
implementation of the so-called PHMC algorithm \cite{PHMC_hitech}.

The above definition of the polynomial $p_{n,\epsilon}(q)$, and 
hence also $P_{n,\epsilon}(s)$, can
straightforwardly be generalized to the case when, instead of the real variable
$q$, a Hermitian matrix is used. 
In this paper we will only study the special matrix $\hat{Q}$ and 
refer to appendix A for its definition.
Here we only mention that $\hat{Q}$ has a band structure with only the diagonal
and a few off-diagonals different from zero. 
We will use for our studies of rounding-error effects
both forms of the polynomial, 
eq.(\ref{pnepsilon}) and eq.(\ref{pol_prod_q_start}). 
We already remark at this point that no qualitative difference 
in our results could be seen using either of the two forms
of the polynomial. 

In order to discuss the rounding errors occurring in the evaluation of
the polynomial $p_{n,\epsilon}$ either in a real variable or in a matrix, 
let us consider the quantities
\bea \label{partial_qprod}
\omega_l(q) & \equiv & | \sqrt{c_l} (q-r_l) \dots
\sqrt{c_1} (q-r_1) q |, \nonumber \\
\Omega_l(\hat{Q}) & \equiv & \| \sqrt{c_l} (\hat{Q}-r_l) \dots
\sqrt{c_1} (\hat{Q}-r_1) \hat{Q}\Phi_{\rm in} \|/\| \Phi_{\rm in}\|, \nonumber \\
 & & l\in \left\{1,2,...,2n+1\right\},\;\;\; r_{2n+1}=0,\;\;\;c_{2n+1}=1\; .
\eea
In eqs.(\ref{partial_qprod}), $\Phi_{\rm in}$ is a suitable 
vector on which the matrices act. The constants $r_{2n+1}$ and 
$c_{2n+1}$ are chosen such 
that the last factor in $\omega_{2n+1}(q)$ and $\Omega_{2n+1}(\hat{Q})$ 
corresponds
to a multiplication by $q$ and $\hat{Q}$, respectively. 

\begin{figure}[t]
\vspace{-0mm}
\centerline{ \epsfysize=12.5cm
             \epsfxsize=12.5cm
             \epsfbox{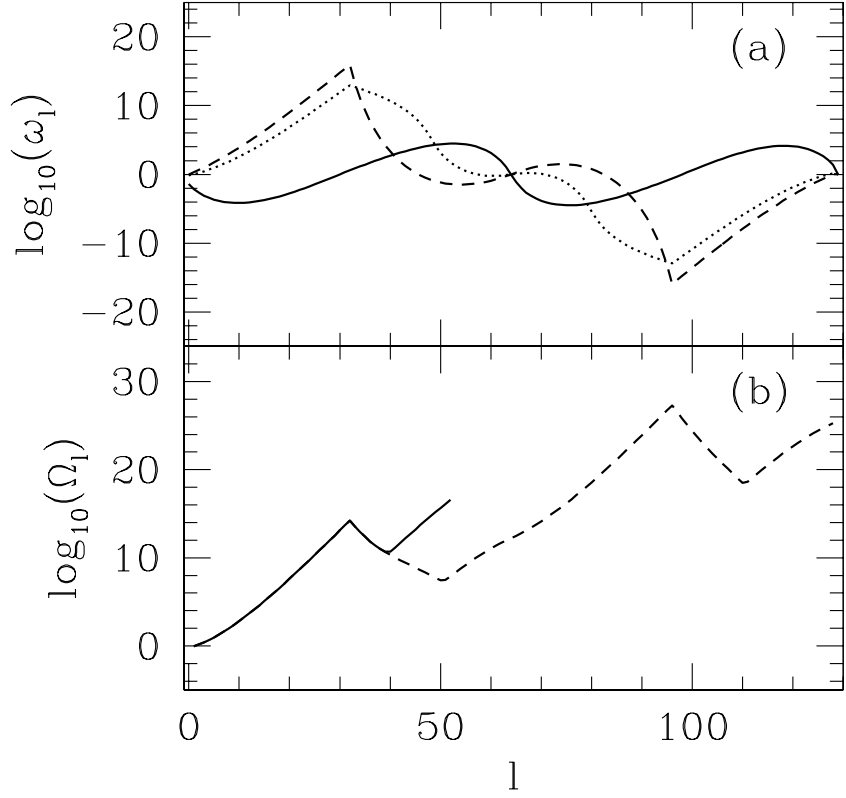}}
\begin{center}
\parbox{12.5cm}{\caption{ \label{fig:naiv_q_story}
The behaviour of $\omega_{l}(q)$
in eq.(\ref{partial_qprod})
as a function of $l$ (a).
The three curves correspond to
$q^2 = \epsilon$ (full line), $q^2 = (1+\epsilon)/2$ (dotted line) and
$q^2 = 1$ (dashed line). In (b) we give
$\Omega_{l}(\hat{Q})$ of eq.(\ref{partial_qprod}) as a function of $l$,
by computing $\Omega_{l}(\hat{Q})$ once with 32-bit (full line) and once with 64-bit
(dashed line) arithmetics.
}}
\end{center}
\end{figure}

In practice 
the values of 
subsequent products, such as $\omega_l(q)$
and $\omega_{l+1}(q)$, may differ substantially. 
This can be clearly seen in Fig.~\ref{fig:naiv_q_story}a.
For the figure we have chosen $n=64$ and $\epsilon=0.0015$,
leading to a relative fit accuracy of $\delta \simeq 0.013$. 
The values of $n$ and $\epsilon$ are motivated by the values 
used in practical applications
such as simulations of lattice QCD. 


We plot $\omega_l(q)$ for values of $q$ at the low end of the interval,
$q^2 = \epsilon$, the middle of the interval, $q^2 = (1+\epsilon)/2$, and the
upper end of the interval, $q^2 = 1$. We will restrict ourselves 
to positive values of $q$. Using the roots as given in eq.(\ref{r_k})
and taking the factors $\sqrt{c_k}$ all identical,
this restriction induces no loss of generality because of the relation 
$\omega_l(q) \omega_{n}(-q) = q \omega_{n+l}(-q) $, $l \in [1,n]$
and $q\neq 0$. 
In the case considered in Fig.~\ref{fig:naiv_q_story}a the oscillations
of $\omega_l(q)$ do not lead to numerical overflows.
Consequently, the final value $\omega_{2n+1}(q)$ comes out to be 
close to $1$, with a deviation consistent with the value of the accuracy parameter
$\delta$. 


The observed behaviour of $\omega_l(q)$ leads us to expect that large precision
losses may affect the evaluation of the same polynomial in the matrix $\hat{Q}$,
when using a product representation, as we do for the computation of 
$\Omega_l(\hat{Q})$, eq.(\ref{partial_qprod}). The $l$-th step of this
computation, yielding $\Phi_{l}$ as a result, amounts
to the multiplication of the vector
\begin{equation}
\Phi_{l-1} = \sqrt{c_{l-1}} (\hat{Q}-r_{l-1}) \cdot \dots \cdot \sqrt{c_1} 
(\hat{Q}-r_1) \hat{Q}\Phi_{\rm in}
\end{equation}
by the matrix $\sqrt{c_{l}} (\hat{Q}-r_{l})$. In order to understand
the relation between the quantities $\omega_j(q)$ and the rounding errors on 
$\Omega_j(\hat{Q})= \| \Phi_j \| / \| \Phi_{\rm in}\| $, it is useful
to think of the vectors $\Phi_j$ and $\Phi_{\rm in}$ as linear combinations
of the eigenvectors of $\hat{Q}$:
\begin{equation} \nn
\Phi_j = \sum_b \langle q_b | \Phi_{\rm in} \rangle
\sqrt{c_j} (q_b - r_j) \cdot \dots \cdot \sqrt{c_1} (q_b-r_1) | q_b \rangle \; ,
\quad \quad (j=1,2, \dots ,2n+1) \; ,
\end{equation} 
where $\hat{Q} | q_b \rangle = q_b | q_b \rangle $ and $\Phi_{\rm in}$ is assumed
to have projections generally non vanishing on all the eigenvectors of $\hat{Q}$.
Let us now go back to situations like the one in Fig.~\ref{fig:naiv_q_story}a
where, for several integer values of $l$, the quantities $\omega_l(q)$ in
a given range of values of $q$ turn
out to be much larger than for all other values of $q$ and they also change substantially
as a function of $l$. It is clear that in such situations the quantities 
$\Omega_l(\hat{Q})$ must substantially change with $l$, too. In particular the situation
$\Omega_{l}(\hat{Q}) \ll \Omega_{l-1}(\hat{Q})$ must occur for some values of $l$,
if the correct final value $\Omega_{2n+1}(\hat{Q}) \simeq \| \Phi_{\rm in}\|$ is 
going to be obtained. But such a situation can only be the result of 
substantial cancellations in the multiplication leading from $\Phi_{l-1}$ to
$\Phi_l$: it is here that we expect the occurrence of large rounding errors,
which then propagate through the whole computation. 
As we will see later, these anticipated precision losses actually occur when
the polynomial in a matrix is evaluated through its product representation. 

The problem itself suggests, however, its solution: the monomial factors
in eq.(\ref{pnepsilon}) or eq.(\ref{pol_prod_q_start}) 
should be ordered, if possible, in such a way that the absolute values 
of all subsequent products of monomials in eq.(\ref{pnepsilon}) or eq.(\ref{pol_prod_q_start})
have the same order of magnitude. 
Of course, whenever possible, evaluating
a polynomial in the product representation should be avoided, because in general 
numerically stable recursion
relations \cite{foxy} or other numerical recipes \cite{numrec} are available. 
However, for some cases, as discussed in the introduction, one has to rely 
on the product representation of the polynomial. 

\section{Ordering schemes}

In this section we want to introduce the different ordering schemes,
that we use
for the monomials in eqs.(\ref{pnepsilon}) and (\ref{pol_prod_q_start}), 
and the well-known, numerically stable Chebyshev method
 for the evaluation of general polynomials.
Throughout this paper we will use the homogeneous distribution
\begin{eqnarray}
c_k = (C_{\rm tot})^{1 \over n}
\nn
\end{eqnarray}
for the normalization constants. 
We remark that 
in principle one could try, at least for the case when a polynomial in 
a matrix is considered, 
to also distribute the normalization constants
$c_k$ in a $k$-dependent way 
to reduce rounding errors. 
However, as expected, 
we only found a very weak dependence of the rounding
errors in the matrix case on the distribution of the $c_k$'s.

\subsection{Definition of ordering schemes}

We start by defining ordering schemes for the monomial factors
in eq.(\ref{pnepsilon}), or equivalently the roots $z_k$ of 
eq.(\ref{roots}). 

\vspace{0.3cm} 
\noindent {\bf Naive ordering}

\begin{figure}[t]
\vspace{-0mm}
\centerline{ \epsfysize=17.5cm
             \epsfxsize=17.5cm
             \epsfbox{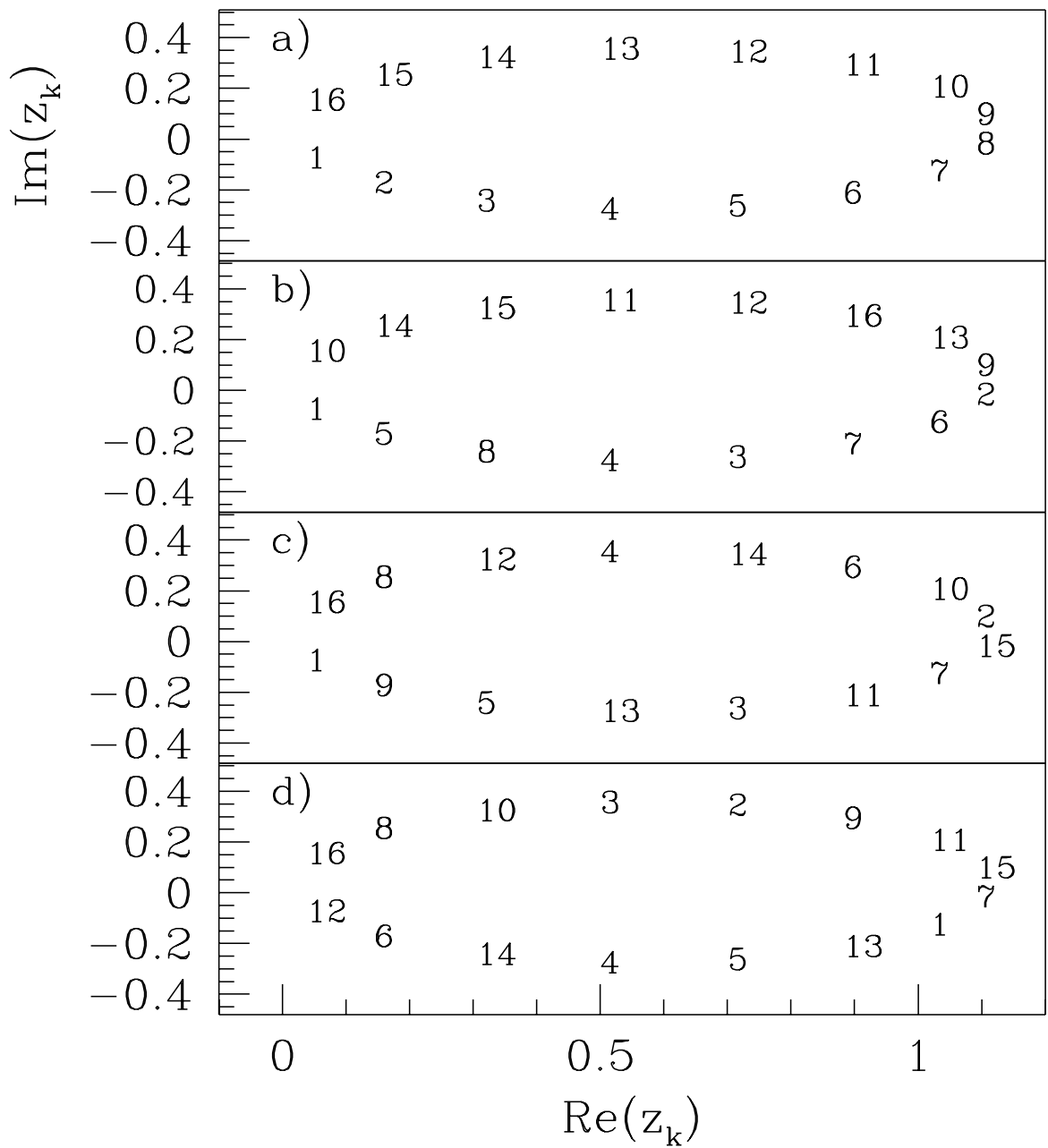}}
\begin{center}
\parbox{12.5cm}{\caption{ \label{fig:index}
The roots $z_k$, eq.~(\ref{roots}), with $k=1,\dots, 16$
and $\epsilon =0.1$ are shown in the complex plane.
Labels of roots indicate in which order they
are used for the numerical evaluation of the
polynomial, eq.~(\ref{pnepsilon}),
within each ordering scheme.
We show in a) the naive, b) the pairing, c)
the bit reversal and d) the Montvay's ordering scheme.
}}
\end{center}
\end{figure}

As naive we regard the ordering 
given by
\begin{equation} \label{znaive}
z^{\rm naive}_k
=
z_k\;, \quad \; k=1,\cdots,n \; ,
\end{equation}
where the roots $z_k$, given in eq.~(\ref{roots}), 
lie on an ellipse in the complex plane. 
In the naive ordering the roots are  
selected from this ellipse by 
starting at the origin and
moving anti-clockwise.
This is indicated in Fig.~\ref{fig:index}a, where the roots are shown
labelled according to the order in which they are used in the evaluation of
the polynomial of eq.~(\ref{pnepsilon}). 
Adopting this ordering of the roots for the construction
of the polynomial in a matrix according to its product representation
gives rise to substantial rounding-error effects.




\vspace{0.3cm}
\noindent {\bf Pairing scheme}

A first improvement over the naive ordering is to 
use a simple
pairing scheme, which amounts to reordering the roots as follows:  
\begin{eqnarray}
z_k^{\rm pair}
=
z_{j(k)}^{\rm naive}\; , \quad \; k=1,\dots,n\; .
\nn
\end{eqnarray}
Let us give the 
reordering index $j(k)$ for the example of
$n$ being a multiple of $8$ and $n'=n/8$.
In the lower half-plane, ${\rm Im}\; z_k < 0$, 
the pairing scheme is achieved by 
\begin{eqnarray}
j 
&=& 
\Bigl\{
1, {n\over 2}, {n\over 4}+1, {n \over 4},
\nn \\&& \phantom{\Bigl\{}
2, {n\over 2}-1, {n\over 4}+2, {n \over 4}-1,
\nn \\&& \phantom{\Bigl\{}
\dots \nn \\
&& \phantom{\Bigl\{}
 n',{n\over 2}-n'+1, {n\over 4}+n', {n \over 4}-n'+1 \Bigr\}
\end{eqnarray}
and for ${\rm Im}\; z_k > 0$  
correspondingly.
An illustration of the ordering in the pairing scheme
is shown in Fig.~\ref{fig:index}b. 

In the case where $n/2$ is not divisible by
4, we search for the next integer $m$, which is smaller than $n/2$ 
and divisible by 4. We then repeat the above described procedure on these 
$m$ roots and simply multiply the remaining roots $z_{m+1} \cdots z_{n/2}$ 
at the end. 

\pagebreak
\vspace{0.3cm}
\noindent {\bf Subpolynomial scheme}

The problem of precision losses in evaluating a polynomial in a matrix
according to its product representation becomes more severe in general when 
increasing the degree $n$ and in the specific case of $P_{n,\epsilon}(s)$,
eq.~(\ref{THEpolynomial}), also when decreasing $\epsilon$.
In order to reduce the effects of rounding errors,
one may therefore be guided by the 
following intuition.
Let us consider the polynomial $P_{n,\epsilon}(s)$ with roots $z_k$ and $n\gg 1$. 
If $m$ is an integer divisor of $n$, the roots $z_{1}, z_{1+m}, z_{1+2m},
\dots, z_{1+(n/m-1)m}$ turn out to be close to the roots
characterising the polynomial
$P_{n',\epsilon}(s)$ of degree $n'=n/m$ 
(note that we keep the same $\epsilon$). Moreover, the 
normalization constants 
$c_{k}= \left( C_{\rm tot} (n) \right)^{1 \over n}$ 
and 
$c'_{k}= \left( C_{\rm tot} (n') \right)^{1 \over n'}$ 
are of the same order (the
dependence on $n$ of $c_k$  turns out to be negligible for large $n$). 
Then 
the product
\begin{equation} \label{u} 
u(s)= \prod_{j=0}^{n/m -1} [c_{j+1}(s-z_{1+jm})] 
\end{equation} 
is a rough approximation of $P_{n',\epsilon}(s)$, $|u(s)-P_{n',\epsilon}(s)| < 1$
for all $\epsilon \le s \le1$. 
The same argument
may be repeated for the other similar sequences of roots, like
$z_{2}, z_{2+m}, z_{2+2m}, \dots z_{2+(n/m-1)m}$, \dots ,
$z_{m}, z_{2m}, z_{3m}, \dots z_{n}$. 

This means that the product eq.(\ref{pnepsilon})
may be split into a product of $m$ subpolynomials, in such a way that each of
them roughly approximates a
polynomial $P_{n',\epsilon}(s)$ of {\it lower} degree $n'=n/m$.
Because of the lower degree of the subpolynomials given by products
such as 
eq.~(\ref{u}), one
may expect that only small changes in the
magnitude of the partial products occur in the intermediate
steps of the evaluation of each of these subpolynomials. 

The reordering of the subpolynomial scheme 

\begin{eqnarray}
z_k^{\rm sp}
=
z_{j(k)}^{\rm naive}
\; , \quad \; k=1,\dots,n 
\nn
\end{eqnarray}
can be represented by 
\begin{eqnarray}
j 
&=& 
\left\{
1, 1+m, 2+2m, \dots, 1 + \left({n\over m}-1\right)m,
\right.
\nn \\&& 
2, 2+m, 2+2m, \dots, 2 + \left({n\over m}-1\right)m,
\nn \\&&  
\dots 
\nn \\&& \left.
m, m+m, m+2m, \dots, m + \left({n\over m}-1\right)m \right\}\; ,
\end{eqnarray}
where $m$
is an integer divisor of $n$.

We found that
$m$ has to be chosen as $m \approx \sqrt{n}$
in order to minimize the changes in the magnitude of the 
partial products occurring in the intermediate steps of the
construction of $P_{n, \epsilon}(s)$.
We remark that the naive ordering 
is reproduced by the two extreme
 choices $m=1$ and
$m=n$. 

\vspace{0.3cm}
\noindent {\bf Bit-reversal scheme}

The subpolynomial scheme can be generalized, leading to what
we will call the bit-reversal scheme. To illustrate how this
scheme works, let us assume that the degree $n$ of the polynomial
is a power of $2$. 
One now starts with the $n$  
monomial factors in eq.~(\ref{pnepsilon}), chooses $m=n/2$ and
applies the subpolynomial scheme resulting in $m$ binomial
factors. We then proceed to choose an $m'=m/2$ and again applying
the subpolynomial scheme to these $m$ binomial factors which leaves us 
with $m'$ subpolynomials each of degree $4$. The procedure can be iterated
until we are left with only one subpolynomial having the degree of
the polynomial itself. 
The above sketched procedure can be realized in practice by 
first representing the integer label (counting from $0$ to $n-1$) 
of the roots in the naive order
by its bit representation. The desired order is then obtained
by simply reversing the bits in this representation. 
The resulting reordering of the roots is shown in
Fig.~\ref{fig:index}c, with $n=16$ as an example. 

For $n$ not a power of 2,
we pad with dummy roots, chosen to be zero for instance, until the 
artificial  
number of roots is a power of 2.
The bit-reversal 
procedure can then be applied as described above.
Afterwards, the dummy roots have to be eliminated from the sequence.

\vspace{0.3cm}
\noindent{\bf Montvay's scheme}

Recently, Montvay \cite{mon97}
suggested to order the roots according to an    
optimization procedure that can be implemented numerically.       
Let us shortly sketch how Montvay's ordering scheme works and
refer to \cite{mon97} for further details. Let us assume that we have 
already the optimized order of
the roots for the partial product $P_{n,\epsilon}^{l}(s)$, eq.~(\ref{plepsilon}). 
Then the values of 
$|sP_{n,\epsilon}^{l}(s)(s-z)|$ are computed for all $z$ taken from 
the set of roots not already used.
The values of $s$ are 
taken from a large enough discrete set of points, 
 $\left\{s_1,\dots,s_N\right\}$, which are all
in the interval $\left[\epsilon,1\right]$.               
Now, the maximal ratio over $s\in \left\{s_1,\dots,s_N\right\}$
of all values 
$|sP_{n,\epsilon}^{l}(s)(s-z)|$, i.e.
\[
 \max_{ s \in \left\{s_1,\dots,s_N\right\} } |s P_{n,\epsilon}^{l} (s) (s-z)|
/
 \min_{ s \in \left\{s_1,\dots,s_N\right\} } |s P_{n,\epsilon}^{l} (s) (s-z)|
\; ,  
\]
 is computed for each root $z$ separately. 
Finally that root is taken which gives the {\em lowest} of
these maximal ratios. Starting with the trivial polynomial 
$P_{n,\epsilon}^{0}(s)=1$, this procedure obviously 
defines a scheme according to which the roots can be ordered iteratively. 
We show in Fig.~\ref{fig:index}d the resulting order of the roots using Montvay's scheme
by again labelling the roots in the order in which they are used to compute the 
polynomial of eq.~(\ref{pnepsilon}). 
It is clear that
Montvay's ordering scheme implements by construction our intuitive criterion that
the changes in the magnitude of subsequent partial products should be minimized.


\vspace{0.3cm}
\noindent{\bf Ordering schemes for the roots $\left\{r_j\right\}$ }

The above discussion concerned the ordering of monomial factors in eq.(\ref{pnepsilon}). 
If one wants to use the product representation,
eq.(\ref{pol_prod_q_start}), the ordering of the monomial factors 
can be obtained by first ordering the roots $z_k$ in one of the
ways described above and then defining the $2n$ roots $r_j$ as 
in eq.(\ref{r_k}). 
The one exception is the case of Montvay's scheme, where the $r_j$'s 
themselves have to be ordered,  
in full analogy with the procedure desribed above
for the case of the product representation (\ref{pnepsilon}).  We remark that the ordering
schemes for  $\left\{r_j \; : \quad j=1, \dots , 2n \right\}$ which we consider in the
present paper, always satisfy the relation in the second line of eq.(\ref{r_k}). Although
there is in principle much more freedom, this restriction turns out to be highly convenient for
the application of Monte Carlo simulation algorithms.

\subsection{Clenshaw recursion} 

Any polynomial ${\mathcal P}_m(u)$ of degree $m$ in the real variable $u$, with $u \in [0,1]$,
can be expressed as a linear combination of Chebyshev polynomials $T_k^{*}(u)$ 
\[ {\mathcal P}_m(u) = \frac{1}{2} a_0 T_0^{*}(u) +
\sum_{k=1}^{m} a_k T_k^{*}(u)  \]
where $a_0, a_1 \dots , a_m$ are suitable coefficients 
and the definition of $T_k^{*}(u)$ is given by 
eq.(\ref{T_r}) and $\cos\theta = 2u-1$.
 
A way \cite{foxy,numrec} of computing ${\mathcal P}_m$ is to use the formula
\be \label{clenshaw_end}
{\mathcal P}_m(u) = \frac{1}{2} (b_0(u) - b_2(u))\; ,
\ee
where the right-hand side has to be evaluated through the recursion
relation:
\begin{eqnarray} \label{clenshaw}
b_{m+2}(u) & = & b_{m+1}(u) = 0  \nonumber \\
b_k(u) \quad & = & 2(2u-1)b_{k+1}(u) - b_{k+2}(u) + a_k \; ,
\end{eqnarray}
for all integers $k$ starting from $m$ down to $0$. It is also possible to prove
that the total rounding error on the final result $(b_0(u) - b_2(u))/2$
cannot exceed the arithmetic sum of the rounding errors occurring
in each step of (\ref{clenshaw}).
 
This well-known, numerically stable method \cite{foxy},\cite{numrec} can
be applied to the evaluation of $P_{n,\epsilon}(s)$ as well as of
$s P_{n,\epsilon}(s)$. In the latter case, if we consider 
$s P_{n,\epsilon}(s)$ as a polynomial of degree $n+1$ in
$u=(s-\epsilon)(1-\epsilon)$, its expression as a linear
combination of Chebyshev polynomials in $u$ can be read
from eq.(\ref{THEpolynomial}):
\[ s P_{n,\epsilon}(s) = T_0^{*}(u) + \rho T_{n+1}^{*}(u)  \; .\]
We can then evaluate $s P_{n,\epsilon}(s)$ through the Clenshaw relation
(\ref{clenshaw}) by taking $a_0=2$, $a_{n+1}=\rho$ and all the other
$a_k$'s vanishing. 
The Clenshaw recursion will serve
us in the following as a reference procedure for the numerical evaluation
of the polynomial $\hat{Q}^2 P_{n,\epsilon}(\hat{Q}^2)$.

\subsection{Ordering schemes at work: a first look} 

As discussed in section 2, the large oscillations of  
$\omega_l(q)$, eq.~(\ref{partial_qprod}), can easily lead to precision losses
when constructing the same polynomial in a matrix. 
This can be seen in 
Fig.~\ref{fig:naiv_q_story}b for the case of the matrix $\hat{Q}$
(see Appendix A for the definition of the matrix $\hat{Q}$).
There we have considered
an $8^3\cdot 16$ lattice with gauge group SU(2), 
at $\beta=1.75$, $\kappa = 0.165$ and $c_{\rm sw} =0$. 
We compare $\Omega_l(\hat{Q})$ computed in 32-bit precision
(solid line) 
with the one computed in 64-bit precision (dashed line). 
We remark at this point that, although matrix multiplications
are performed in 32- or 64-precision, scalar products
are always evaluated with 64-bit precision.        
As starting vector $\Phi_{in}$ we use a Gaussian random vector.

The picture confirms our expectations: 
as long as the values of $\Omega_l(\hat{Q})$ are growing with
$l$, both curves are basically
identical. When $\Omega_l(\hat{Q})$ starts to decrease, however, 
the values for $\Omega_l(\hat{Q})$ obtained with 32-bit precision 
deviate strongly from the ones obtained with 64-bit precision.
In fact, instead of decreasing, $\Omega_l(\hat{Q})$ 
computed with the 32-bit precision version of the program, even increases and eventually 
runs into a numerical overflow. 
This is no surprise, of course: as discussed above,
when $\Omega_{l}(\hat{Q}) \ll \Omega_{l-1}(\hat{Q})$, large
cancellations must occur, leading to the observed precision losses.
Note, moreover, 
that even the values for $\Omega_l(\hat{Q})$ obtained with a 64-bit precision 
are affected by large rounding errors: the
final value $\Omega_{2n+1}(\hat{Q})$ is completely wrong, namely 
$O(10^{25})$ instead of $O(1)$, as it should be. 
Clearly, using only 64-bit
precision reduces the precision losses (no overflow is observed),
but it certainly is not sufficient to keep in general rounding errors
on a tolerable level.

\begin{figure}[t]
\vspace{-0mm}
\centerline{ \epsfysize=12.5cm
             \epsfxsize=12.5cm
             \epsfbox{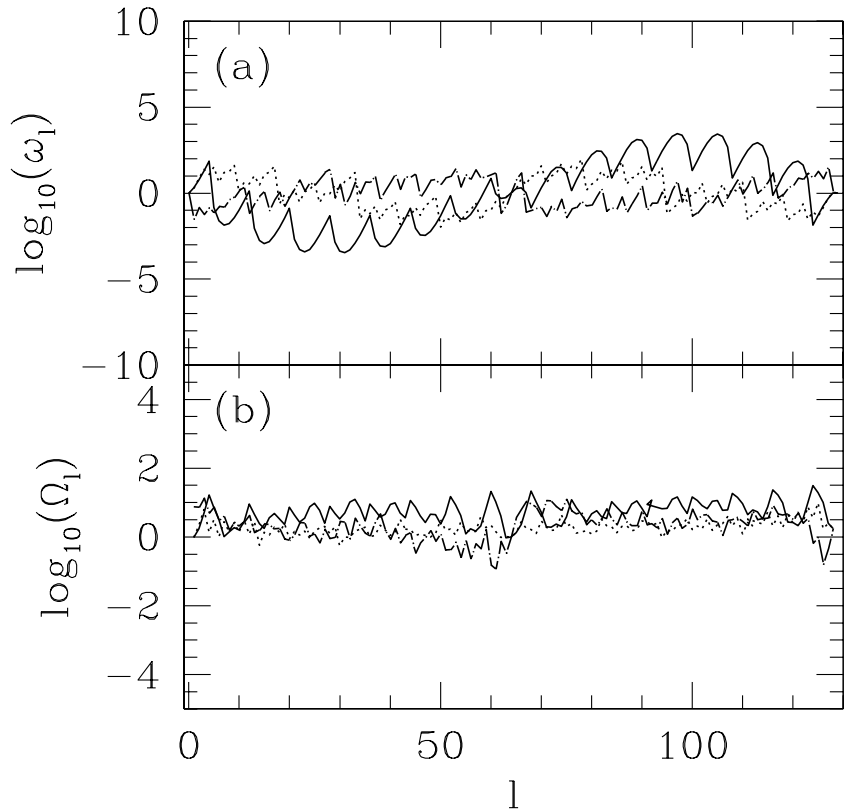}}
\begin{center}
\parbox{12.5cm}{\caption{ \label{fig:orderedroots}
 The behaviour of (a) $\omega_{l}(q)$,
eq.(\ref{partial_qprod}), and (b) of $\Omega_{l}(\hat{Q})$, eq.(\ref{partial_qprod}),
as a function of $l$.
We use the subpolynomial (solid line), the bit reversal
(dotted line) and Montvay's (dash-dotted line) ordering schemes.
For Fig.3a, we have chosen only $q=1$.
All results shown in the figure are obtained using 32-bit precision.
}}
\end{center}
\end{figure}

Figure~\ref{fig:orderedroots} shows how the different, improved ordering schemes help. 
In Fig.\ref{fig:orderedroots}a we plot $\omega_l(q)$ for three ordering schemes,
the subpolynomial (solid line), the bit-reversal (dashed line) and Montvay's
scheme (dash-dotted line). We only show the curves for $q=1$, i.e. the worst case
in Fig.\ref{fig:naiv_q_story}a. For other
values of $q$ the picture looks very similar. As compared with
Fig.~\ref{fig:naiv_q_story}a, the large oscillations  
are strongly suppressed.
As a consequence, when now $\Omega_l(\hat{Q})$ is constructed with
32-bit precision, according to 
the improved ordering schemes, numerical overflows are avoided
and the desired fit accuracy is reached, as demonstrated in 
Fig.\ref{fig:orderedroots}b.

\section{Quantitative tests}

After the qualitative tests of the ordering schemes discussed in the previous section,
we would now like to turn to more quantitative results. 
Guided by the observation that 
the evaluation of a polynomial
in a single variable gives information on the precision losses
that may occur 
when evaluating 
the same polynomial in a matrix, we will first investigate single variable  
estimators for rounding errors.
Then we will discuss the rounding-error effects that arise in the
numerical construction of the polynomial $\hat{Q}^2 P_{n,\epsilon}(\hat{Q}^2)$. 
The results of this section only
refer to ordering schemes for polynomials in the product representation
of eq.(\ref{pnepsilon}). We stress that there would be no qualitative
difference in our results if the representation 
eq.(\ref{pol_prod_q_start}) had been taken. 

\subsection{Estimators of rounding errors for a single variable}

A possible way of defining single variable estimators  
for rounding-error effects consists in quantifying
the magnitude of the oscillations of 
$\omega_l(q)$. As a first step, let us evaluate        
for a given $l$ the maximal and the minimal value 
of $|P_{n,\epsilon}^{l}(s)|$, eq.(\ref{plepsilon}), 
over the interval $0\le s \le 1$. 
The ratio of the maximal to the minimal value,
i.e. $\tilde{R}_l=\max_{s \in [0,1]} |P_{n,\epsilon}^{l} (s)|/
\min_{s \in [0,1]} |P_{n,\epsilon}^{l} (s)|$, is then a measure
of how different the order of magnitude of the
$l$-th partial product can be for different values of $s$.
Building the maximum of $\tilde{R}_l$ with respect to $l$, 
we get a quantity independent of $l$ and $s$:
\begin{equation} \label{rmax} 
R_{\rm max} 
=
\max_{ l \in \{1, \dots, n\} } \left\{
{ \max_{s \in [0,1]} |P_{n,\epsilon}^{l} (s)|
\over
 \min_{s \in [0,1]} |P_{n,\epsilon}^{l} (s)|}
\right\}\; .
\end{equation}
It is clear that 
$R_{\rm max}$  
has to be smaller
than the largest representable number on a given computer 
to guarantee the stability of the evaluation of the full polynomial.  
This is actually sufficient to exclude the occurrence of overflows
or underflows in the evaluation of the considered partial products. 
If $R_{\rm max}$ fulfils this condition but still assumes
large values, in the case of a polynomial in a matrix
a reliable numerical result cannot be expected, since rounding
errors are likely to lead to a substantial loss of precision.
Moreover, even if $R_{\rm max}$ is reasonably small, it is still
possible that the quantity
$\max_{s \in [0,1]} |P_{n,\epsilon}^{l} (s)|$
shows large oscillations as a
function of $l$. As a consequence, another quantity of interest is the
maximum value of the partial products itself: 
\begin{equation} \label{M}
M_{\rm max}
=
\max_{s \in [0,1], l \in \{1, \dots, n\} } |P_{n,\epsilon}^{l}(s)|\;.
\end{equation}

This again has to  
be smaller 
than the largest representable number in order not to run into
overflow. 
Note that $R_{\rm max}$ and $M_{\rm max}$
are computed for $s \in [0,1]$, whereas the 
polynomial in eq.(\ref{pnepsilon}) has a given relative fit accuracy only in the
interval $s \in [\epsilon,1]$. 
However, 
as will be explicitly demonstrated below, our results for 
$R_{\rm max}$ and $M_{\rm max}$ do not depend very much on the
choice of the lower end of the interval. 

A final remark is that the values of $R_{\rm max}$ and $M_{\rm max}$ should be 
taken only as a first indication for the size of the rounding errors.
Since it is not guaranteed that in a practical case, for the
polynomial in a matrix, the value of $R_{\rm max}$ or $M_{\rm max}$ is actually assumed, 
it is very possible that 
$R_{\rm max}$ and $M_{\rm max}$ 
may overestimate the rounding errors. 
On the other hand, since in the case of a polynomial in a matrix
the rounding errors occurring in different intermediate steps
of the numerical computation can easily accumulate, 
$R_{\rm max}$ and $M_{\rm max}$ might also yield an underestimate. 

In order to compute the values of $R_{\rm max}$ and $M_{\rm max}$ 
we take $5000$ values of $s$, equally spaced in the
interval $[0,1]$. 
We explicitly checked that the values of $R_{\rm max}$ do not depend
very much on the lower end of  the interval $[s_{\rm min},1]$
from which $s$ is taken.
In Fig.~\ref{fig:scale}
we show  
$R_{\rm max}$, in the case of the bit-reversal scheme, 
as a function of the
lower end of the interval  $s_{\rm min}$, measured in units of 
the parameter $\epsilon$.
The data refer to polynomials of different degree,
$n=30$, $n=86$ and $n=146$, with the parameter $\epsilon$
determined by the fixed relative fit accuracy $\delta=0.001$.
As Fig.~\ref{fig:scale} shows, the dependence of $R_{\rm max}$ on 
$s_{\rm min}$ is very weak. For the other ordering schemes, we find
a similar behaviour of $R_{\rm max}$ as a function of $s_{\rm min}$.
This justifies the use of $s_{\rm min}=0$ that we have adopted 
for the numerical tests described below. 

We start by comparing the subpolynomial and the bit-reversal schemes, as they
are closely related to each other. In Fig.~\ref{fig:d0.1} we show $R_{\rm max}$ and $M_{\rm max}$
as a function of the degree $n$ of the polynomial
$P_{n,\epsilon}(s)$ of eq.(\ref{pnepsilon}), keeping the relative fit accuracy
$\delta =0.1$ constant by adjusting the parameter $\epsilon$. 
For the subpolynomial scheme, the divisor $m$ is chosen to be
$m\approx\sqrt{n}$. Figure~\ref{fig:d0.1} clearly confirms our expectation that the bit-reversal 
scheme, considered as a generalization of the subpolynomial ordering scheme,
gives smaller values of $R_{\rm max}$ and $M_{\rm max}$. For degrees of the 
polynomial $n>40$ 
the ``dangerous'' oscillations 
are substantially suppressed in the bit-reversal scheme 
compared with the subpolynomial scheme. 

\begin{figure}[tbp]

\setlength{\unitlength}{0.1bp}
\special{!
/gnudict 40 dict def
gnudict begin
/Color false def
/Solid false def
/gnulinewidth 5.000 def
/vshift -33 def
/dl {10 mul} def
/hpt 31.5 def
/vpt 31.5 def
/M {moveto} bind def
/L {lineto} bind def
/R {rmoveto} bind def
/V {rlineto} bind def
/vpt2 vpt 2 mul def
/hpt2 hpt 2 mul def
/Lshow { currentpoint stroke M
  0 vshift R show } def
/Rshow { currentpoint stroke M
  dup stringwidth pop neg vshift R show } def
/Cshow { currentpoint stroke M
  dup stringwidth pop -2 div vshift R show } def
/DL { Color {setrgbcolor Solid {pop []} if 0 setdash }
 {pop pop pop Solid {pop []} if 0 setdash} ifelse } def
/BL { stroke gnulinewidth 2 mul setlinewidth } def
/AL { stroke gnulinewidth 2 div setlinewidth } def
/PL { stroke gnulinewidth setlinewidth } def
/LTb { BL [] 0 0 0 DL } def
/LTa { AL [1 dl 2 dl] 0 setdash 0 0 0 setrgbcolor } def
/LT0 { PL [] 0 1 0 DL } def
/LT1 { PL [4 dl 2 dl] 0 0 1 DL } def
/LT2 { PL [2 dl 3 dl] 1 0 0 DL } def
/LT3 { PL [1 dl 1.5 dl] 1 0 1 DL } def
/LT4 { PL [5 dl 2 dl 1 dl 2 dl] 0 1 1 DL } def
/LT5 { PL [4 dl 3 dl 1 dl 3 dl] 1 1 0 DL } def
/LT6 { PL [2 dl 2 dl 2 dl 4 dl] 0 0 0 DL } def
/LT7 { PL [2 dl 2 dl 2 dl 2 dl 2 dl 4 dl] 1 0.3 0 DL } def
/LT8 { PL [2 dl 2 dl 2 dl 2 dl 2 dl 2 dl 2 dl 4 dl] 0.5 0.5 0.5 DL } def
/P { stroke [] 0 setdash
  currentlinewidth 2 div sub M
  0 currentlinewidth V stroke } def
/D { stroke [] 0 setdash 2 copy vpt add M
  hpt neg vpt neg V hpt vpt neg V
  hpt vpt V hpt neg vpt V closepath stroke
  P } def
/A { stroke [] 0 setdash vpt sub M 0 vpt2 V
  currentpoint stroke M
  hpt neg vpt neg R hpt2 0 V stroke
  } def
/B { stroke [] 0 setdash 2 copy exch hpt sub exch vpt add M
  0 vpt2 neg V hpt2 0 V 0 vpt2 V
  hpt2 neg 0 V closepath stroke
  P } def
/C { stroke [] 0 setdash exch hpt sub exch vpt add M
  hpt2 vpt2 neg V currentpoint stroke M
  hpt2 neg 0 R hpt2 vpt2 V stroke } def
/T { stroke [] 0 setdash 2 copy vpt 1.12 mul add M
  hpt neg vpt -1.62 mul V
  hpt 2 mul 0 V
  hpt neg vpt 1.62 mul V closepath stroke
  P  } def
/S { 2 copy A C} def
end
}
\begin{picture}(3600,2160)(0,0)
\special{"
gnudict begin
gsave
50 50 translate
0.100 0.100 scale
0 setgray
/Helvetica findfont 100 scalefont setfont
newpath
-500.000000 -500.000000 translate
LTa
LTb
600 269 M
63 0 V
2754 0 R
-63 0 V
600 453 M
63 0 V
2754 0 R
-63 0 V
600 637 M
63 0 V
2754 0 R
-63 0 V
600 1005 M
63 0 V
2754 0 R
-63 0 V
600 1373 M
63 0 V
2754 0 R
-63 0 V
600 1741 M
63 0 V
2754 0 R
-63 0 V
600 2109 M
63 0 V
2754 0 R
-63 0 V
600 251 M
0 63 V
0 1795 R
0 -63 V
1024 251 M
0 31 V
0 1827 R
0 -31 V
1272 251 M
0 31 V
0 1827 R
0 -31 V
1448 251 M
0 31 V
0 1827 R
0 -31 V
1584 251 M
0 31 V
0 1827 R
0 -31 V
1696 251 M
0 31 V
0 1827 R
0 -31 V
1790 251 M
0 31 V
0 1827 R
0 -31 V
1872 251 M
0 31 V
0 1827 R
0 -31 V
1944 251 M
0 31 V
0 1827 R
0 -31 V
2009 251 M
0 63 V
0 1795 R
0 -63 V
2433 251 M
0 31 V
0 1827 R
0 -31 V
2681 251 M
0 31 V
0 1827 R
0 -31 V
2857 251 M
0 31 V
0 1827 R
0 -31 V
2993 251 M
0 31 V
0 1827 R
0 -31 V
3105 251 M
0 31 V
0 1827 R
0 -31 V
3199 251 M
0 31 V
0 1827 R
0 -31 V
3281 251 M
0 31 V
0 1827 R
0 -31 V
3353 251 M
0 31 V
0 1827 R
0 -31 V
3417 251 M
0 63 V
0 1795 R
0 -63 V
600 251 M
2817 0 V
0 1858 V
-2817 0 V
600 251 L
LT0
1568 1864 D
712 439 D
806 475 D
888 510 D
1024 571 D
1136 636 D
1272 729 D
1349 799 D
1478 924 D
1584 1051 D
1696 1210 D
1808 1399 D
1916 1620 D
2021 1873 D
LT1
1568 1764 A
712 390 A
806 400 A
888 409 A
1024 428 A
1136 437 A
1272 452 A
1349 460 A
1478 475 A
1584 491 A
1696 531 A
1808 576 A
1916 628 A
2021 689 A
2130 766 A
2240 857 A
2347 966 A
2456 1099 A
2569 1260 A
2676 1446 A
LT2
1568 1664 B
712 390 B
806 393 B
888 449 B
1024 443 B
1136 465 B
1272 460 B
1349 544 B
1478 505 B
1584 546 B
1696 526 B
1808 595 B
1916 568 B
2021 615 B
2130 588 B
2240 688 B
2347 624 B
2456 655 B
2569 636 B
2676 718 B
LT3
1568 1564 C
712 416 C
806 433 C
888 449 C
1024 470 C
1136 456 C
1272 490 C
1349 493 C
1478 508 C
1584 552 C
1696 551 C
1808 593 C
1916 582 C
2021 635 C
2130 652 C
2240 678 C
2347 730 C
2456 742 C
2569 823 C
2676 908 C
stroke
grestore
end
showpage
}
\put(1448,1564){\makebox(0,0)[r]{Montvay}}
\put(1448,1664){\makebox(0,0)[r]{Bit-reversal}}
\put(1448,1764){\makebox(0,0)[r]{Pairing}}
\put(1448,1864){\makebox(0,0)[r]{Naive}}
\put(2008,51){\makebox(0,0){$n$}}
\put(100,1180){%
\special{ps: gsave currentpoint currentpoint translate
270 rotate neg exch neg exch translate}%
\makebox(0,0)[b]{\shortstack{$R_{\rm max}$}}%
\special{ps: currentpoint grestore moveto}%
}
\put(3417,151){\makebox(0,0){1000}}
\put(2009,151){\makebox(0,0){100}}
\put(600,151){\makebox(0,0){10}}
\put(540,2109){\makebox(0,0)[r]{1e+30}}
\put(540,1741){\makebox(0,0)[r]{1e+24}}
\put(540,1373){\makebox(0,0)[r]{1e+18}}
\put(540,1005){\makebox(0,0)[r]{1e+12}}
\put(540,637){\makebox(0,0)[r]{1e+06}}
\put(540,453){\makebox(0,0)[r]{1000}}
\put(540,269){\makebox(0,0)[r]{1}}
\end{picture}
\setlength{\unitlength}{0.1bp}
\special{!
/gnudict 40 dict def
gnudict begin
/Color false def
/Solid false def
/gnulinewidth 5.000 def
/vshift -33 def
/dl {10 mul} def
/hpt 31.5 def
/vpt 31.5 def
/M {moveto} bind def
/L {lineto} bind def
/R {rmoveto} bind def
/V {rlineto} bind def
/vpt2 vpt 2 mul def
/hpt2 hpt 2 mul def
/Lshow { currentpoint stroke M
  0 vshift R show } def
/Rshow { currentpoint stroke M
  dup stringwidth pop neg vshift R show } def
/Cshow { currentpoint stroke M
  dup stringwidth pop -2 div vshift R show } def
/DL { Color {setrgbcolor Solid {pop []} if 0 setdash }
 {pop pop pop Solid {pop []} if 0 setdash} ifelse } def
/BL { stroke gnulinewidth 2 mul setlinewidth } def
/AL { stroke gnulinewidth 2 div setlinewidth } def
/PL { stroke gnulinewidth setlinewidth } def
/LTb { BL [] 0 0 0 DL } def
/LTa { AL [1 dl 2 dl] 0 setdash 0 0 0 setrgbcolor } def
/LT0 { PL [] 0 1 0 DL } def
/LT1 { PL [4 dl 2 dl] 0 0 1 DL } def
/LT2 { PL [2 dl 3 dl] 1 0 0 DL } def
/LT3 { PL [1 dl 1.5 dl] 1 0 1 DL } def
/LT4 { PL [5 dl 2 dl 1 dl 2 dl] 0 1 1 DL } def
/LT5 { PL [4 dl 3 dl 1 dl 3 dl] 1 1 0 DL } def
/LT6 { PL [2 dl 2 dl 2 dl 4 dl] 0 0 0 DL } def
/LT7 { PL [2 dl 2 dl 2 dl 2 dl 2 dl 4 dl] 1 0.3 0 DL } def
/LT8 { PL [2 dl 2 dl 2 dl 2 dl 2 dl 2 dl 2 dl 4 dl] 0.5 0.5 0.5 DL } def
/P { stroke [] 0 setdash
  currentlinewidth 2 div sub M
  0 currentlinewidth V stroke } def
/D { stroke [] 0 setdash 2 copy vpt add M
  hpt neg vpt neg V hpt vpt neg V
  hpt vpt V hpt neg vpt V closepath stroke
  P } def
/A { stroke [] 0 setdash vpt sub M 0 vpt2 V
  currentpoint stroke M
  hpt neg vpt neg R hpt2 0 V stroke
  } def
/B { stroke [] 0 setdash 2 copy exch hpt sub exch vpt add M
  0 vpt2 neg V hpt2 0 V 0 vpt2 V
  hpt2 neg 0 V closepath stroke
  P } def
/C { stroke [] 0 setdash exch hpt sub exch vpt add M
  hpt2 vpt2 neg V currentpoint stroke M
  hpt2 neg 0 R hpt2 vpt2 V stroke } def
/T { stroke [] 0 setdash 2 copy vpt 1.12 mul add M
  hpt neg vpt -1.62 mul V
  hpt 2 mul 0 V
  hpt neg vpt 1.62 mul V closepath stroke
  P  } def
/S { 2 copy A C} def
end
}
\begin{picture}(3600,2160)(0,0)
\special{"
gnudict begin
gsave
50 50 translate
0.100 0.100 scale
0 setgray
/Helvetica findfont 100 scalefont setfont
newpath
-500.000000 -500.000000 translate
LTa
LTb
600 269 M
63 0 V
2754 0 R
-63 0 V
600 453 M
63 0 V
2754 0 R
-63 0 V
600 637 M
63 0 V
2754 0 R
-63 0 V
600 1005 M
63 0 V
2754 0 R
-63 0 V
600 1373 M
63 0 V
2754 0 R
-63 0 V
600 1741 M
63 0 V
2754 0 R
-63 0 V
600 2109 M
63 0 V
2754 0 R
-63 0 V
600 251 M
0 63 V
0 1795 R
0 -63 V
1024 251 M
0 31 V
0 1827 R
0 -31 V
1272 251 M
0 31 V
0 1827 R
0 -31 V
1448 251 M
0 31 V
0 1827 R
0 -31 V
1584 251 M
0 31 V
0 1827 R
0 -31 V
1696 251 M
0 31 V
0 1827 R
0 -31 V
1790 251 M
0 31 V
0 1827 R
0 -31 V
1872 251 M
0 31 V
0 1827 R
0 -31 V
1944 251 M
0 31 V
0 1827 R
0 -31 V
2009 251 M
0 63 V
0 1795 R
0 -63 V
2433 251 M
0 31 V
0 1827 R
0 -31 V
2681 251 M
0 31 V
0 1827 R
0 -31 V
2857 251 M
0 31 V
0 1827 R
0 -31 V
2993 251 M
0 31 V
0 1827 R
0 -31 V
3105 251 M
0 31 V
0 1827 R
0 -31 V
3199 251 M
0 31 V
0 1827 R
0 -31 V
3281 251 M
0 31 V
0 1827 R
0 -31 V
3353 251 M
0 31 V
0 1827 R
0 -31 V
3417 251 M
0 63 V
0 1795 R
0 -63 V
600 251 M
2817 0 V
0 1858 V
-2817 0 V
600 251 L
LT0
1568 1864 D
712 403 D
806 427 D
888 450 D
1024 493 D
1136 536 D
1272 599 D
1349 647 D
1478 733 D
1584 823 D
1696 932 D
1808 1064 D
1916 1217 D
2021 1393 D
2130 1610 D
2240 1870 D
LT1
1568 1764 A
712 372 A
806 377 A
888 385 A
1024 404 A
1136 404 A
1272 415 A
1349 423 A
1478 432 A
1584 441 A
1696 464 A
1808 459 A
1916 467 A
2021 494 A
2130 536 A
2240 586 A
2347 646 A
2456 727 A
2569 808 A
2676 911 A
LT2
1568 1664 B
712 372 B
806 377 B
888 422 B
1024 410 B
1136 436 B
1272 415 B
1349 508 B
1478 465 B
1584 512 B
1696 464 B
1808 550 B
1916 505 B
2021 555 B
2130 511 B
2240 638 B
2347 536 B
2456 609 B
2569 530 B
2676 655 B
LT3
1568 1564 C
712 393 C
806 408 C
888 422 C
1024 441 C
1136 430 C
1272 460 C
1349 467 C
1478 481 C
1584 515 C
1696 514 C
1808 553 C
1916 551 C
2021 600 C
2130 616 C
2240 588 C
2347 646 C
2456 627 C
2569 669 C
2676 655 C
stroke
grestore
end
showpage
}
\put(1448,1564){\makebox(0,0)[r]{Montvay}}
\put(1448,1664){\makebox(0,0)[r]{Bit-reversal}}
\put(1448,1764){\makebox(0,0)[r]{Pairing}}
\put(1448,1864){\makebox(0,0)[r]{Naive}}
\put(2008,51){\makebox(0,0){$n$}}
\put(100,1180){%
\special{ps: gsave currentpoint currentpoint translate
270 rotate neg exch neg exch translate}%
\makebox(0,0)[b]{\shortstack{$M_{\rm max}$}}%
\special{ps: currentpoint grestore moveto}%
}
\put(3417,151){\makebox(0,0){1000}}
\put(2009,151){\makebox(0,0){100}}
\put(600,151){\makebox(0,0){10}}
\put(540,2109){\makebox(0,0)[r]{1e+30}}
\put(540,1741){\makebox(0,0)[r]{1e+24}}
\put(540,1373){\makebox(0,0)[r]{1e+18}}
\put(540,1005){\makebox(0,0)[r]{1e+12}}
\put(540,637){\makebox(0,0)[r]{1e+06}}
\put(540,453){\makebox(0,0)[r]{1000}}
\put(540,269){\makebox(0,0)[r]{1}}
\end{picture}
\caption{\label{fig:d0.001}
$R_{\rm max}$, eq.~(\ref{rmax}), 
and $M_{\rm max}$, eq.~(\ref{M}),
are shown as a function of the
degree of the polynomial
at fixed relative fit accuracy
$\delta=0.001$. 
We compare naive, pairing, bit reversal and Montvay's
ordering schemes.}
\end{figure}
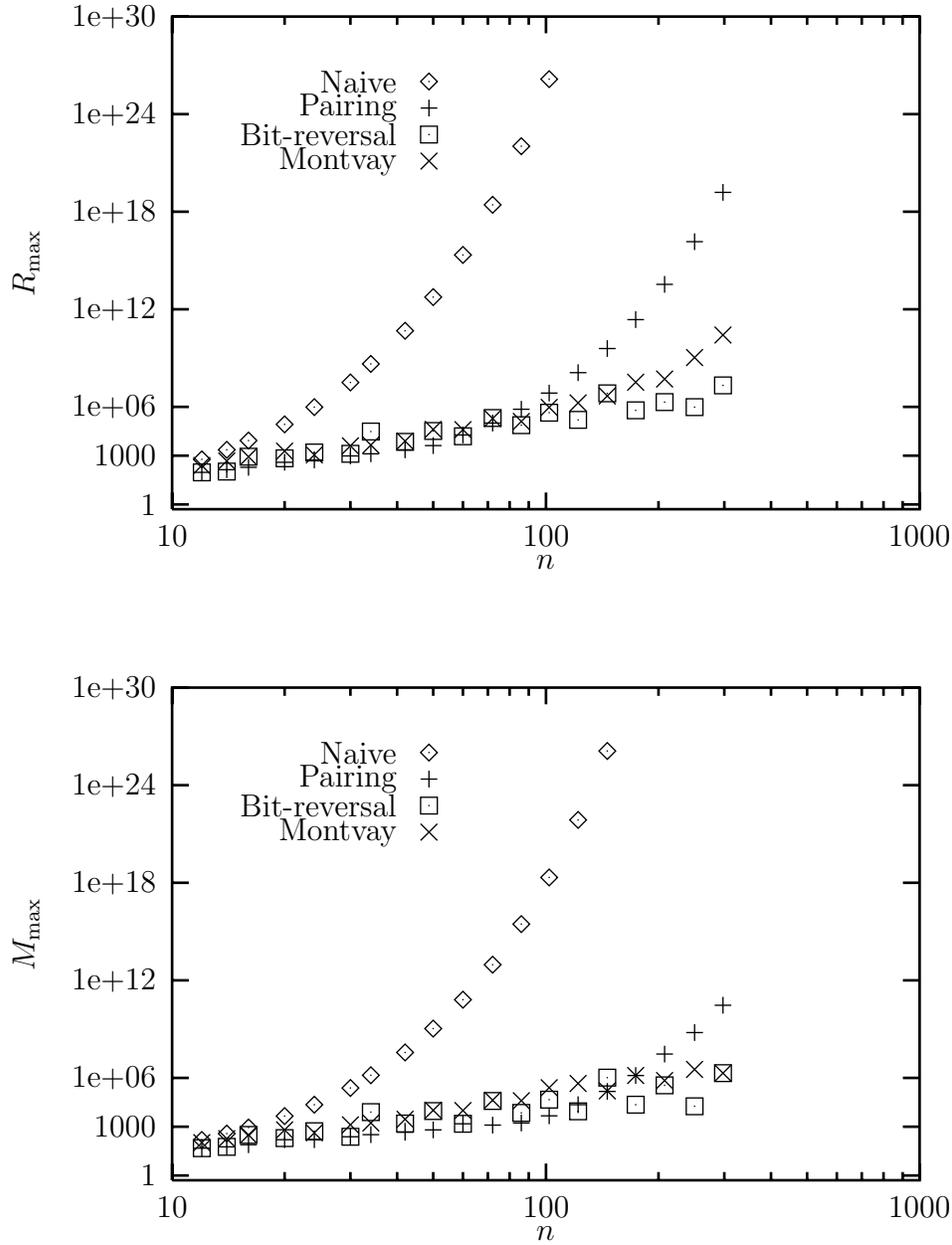

In Fig.~\ref{fig:d0.001} we show the values of $R_{\rm max}$ 
and $M_{\rm max}$ for the bit reversal, the naive, 
the pairing and the Montvay's scheme, at a fixed value 
of $\delta =0.001$,
as a function of $n$. 
Numerical tests for the different ordering schemes were also performed 
at values of $\delta = 0.1$ and $\delta= 0.01$, yielding  
a very similar qualitative behaviour of 
$R_{\rm max}$ and $M_{\rm max}$ as a function of $n$. 

The first striking observation in Fig.~\ref{fig:d0.001} is
that one obtains with the naive ordering, already for 
moderate degrees $n\approx 30$ of the polynomial, large values of $R_{\rm max}$ 
and $M_{\rm max}$; this indicates that very large oscillations
of the partial products occur in the intermediate steps
of the construction of the polynomial.
Using the naive scheme on 
machines with 32-bit precision
or even with 64-bit precision, a safe evaluation of the polynomial
in the product representation can certainly not be guaranteed. 

The behaviour of the values of $R_{\rm max}$ and $M_{\rm max}$ 
obtained by using 
the naive ordering scheme clearly 
demonstrates the necessity of finding better ordering schemes.
That such ordering schemes 
do exist is also demonstrated in Fig.~\ref{fig:d0.001}. 
For $n < 100$, the values for 
$R_{\rm max}$ and $M_{\rm max}$ obtained from 
the pairing, bit reversal and Montvay's 
schemes are close to each other and many orders of magnitude below the ones
of the naive scheme. However, for $n>120$, the values of 
$R_{\rm max}$ from these ordering schemes also start to 
deviate from each other. 
Taking the values of $R_{\rm max}$ and $M_{\rm max}$ as an estimate of the effects 
of rounding errors arising in the construction of polynomials in matrix,
it seems that the bit reversal and Montvay's scheme are the most effective, in reducing
these effects, out of the ordering schemes investigated here. 


\subsection{Quantitative tests for a polynomial in a matrix}

The numerical tests involving a polynomial 
in the matrix $\hat{Q}$, eq.(\ref{preq}),
are performed using a sample of  
thermalized SU(3) gauge field configurations 
on $8^3\cdot 16$ lattices. All numerical computations
were done on 
the massively parallel Alenia Quadrics (APE) machines, 
which have only 32-bit precision. 
Simulation parameters are chosen to be $\beta=6.8$, 
$\kappa = 0.1343$ and $c_{\rm sw} = 1.42511$. They correspond 
to realistic parameter values as actually used in  simulations
to determine values of $c_{\rm sw}$ non-perturbatively \cite{jansom}. 
Throughout this section we will use Schr\"odinger functional
boundary conditions as mentioned in Appendix A. 
Averaging over the gauge field configurations, for the above choice 
of parameters and $c_M=0.735$, 
the lowest eigenvalue of $\hat{Q}^2$ is $\lambda_{\rm min} = 0.00114(4)$
and the largest is $\lambda_{\rm max} = 0.8721(3)$.
Investigations are performed at values of $(n,\epsilon)$ of    
$(16,0.003)$, $(32,0.003)$, $(64,0.0022)$ and $(100,0.0022)$.
At each of these values of $(n,\epsilon)$ we have 
generated ${\rm O}(50)$ gauge field configurations.
Governed by heuristic arguments \cite{frezzi}, for a real simulation, $\epsilon$ 
should be chosen as   
$\epsilon \approx 2\lambda_{\rm min}$ and $\delta \approx 0.01$, 
which roughly corresponds to the choice $(n,\epsilon)=(64,0.0022)$.          

We apply the matrix $\hat{Q}^2P_{n,\epsilon}(\hat{Q}^2)$, which should be
close to the unit matrix for our choices of $n$ and $\epsilon$, to a 
random Gaussian vector $R_G$ and construct the vectors
\begin{equation} \label{clenvector}
\Phi_{\rm order} = \hat{Q}^2P_{n,\epsilon}(\hat{Q}^2) R_G\; ,
\end{equation} 
where $P_{n,\epsilon}(\hat{Q}^2)$ is defined analogously to
eq.(\ref{pnepsilon}) and evaluated using different 
ordering schemes. The subscript ``order'' stands for naive, pairing, bit reversal, Montvay
and Clenshaw. In the latter case, the polynomial $\hat{Q}^2 P_{n,\epsilon}(\hat{Q}^2)$
is, of course, constructed by using the Clenshaw recursion relation, 
as explained in section 3.2. Following this prescription, one first constructs
the Chebyshev polynomial $T_{n+1}^{*}$, eq.(\ref{T_r}),
with $s$ replaced in the obvious way by the matrix $\hat{Q}^2$.
Since the polynomial we are finally interested in
is given by $\hat{Q}^2P_{n,\epsilon}(\hat{Q}^2)= 1 +\rho T_{n+1}^{*}$, 
see eq.(\ref{THEpolynomial}),
any rounding error
that is induced in the construction of $T_{n+1}^{*}$ is suppressed by
a factor of ${\rm O}(\delta)$ since $| \rho | \le \delta$. 

On a given gauge field configuration and for a given $R_G$ we compute
\begin{equation} \label{Delta}
\Delta = \frac{1}{\sqrt{N}}\| \Phi_{\rm order} - \Phi_{\rm Clenshaw} \|\; ,
\end{equation} 
where $N$ is the number of degrees of freedom of the vector
$\Phi$ and $\|.\|$ denotes the square root vector norm. 

\begin{figure}[t]

\vspace{-0mm}
\centerline{ \epsfysize=13.5cm
             \epsfxsize=13.5cm
             \epsfbox{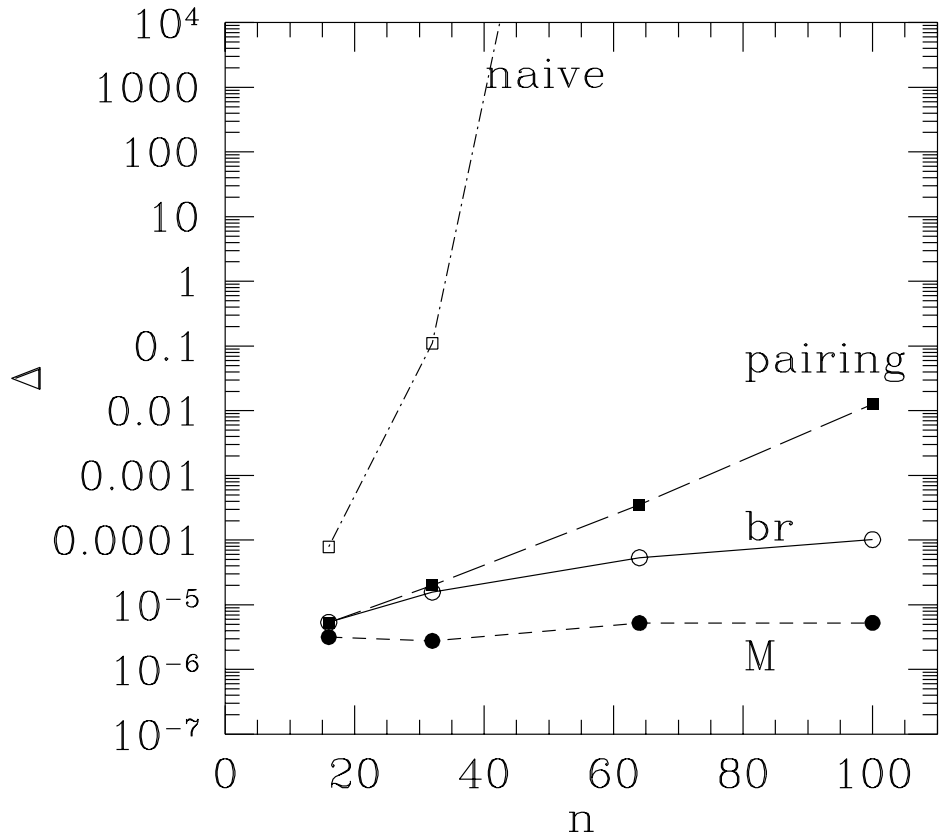}}
\begin{center}
\parbox{12.5cm}{\caption{ \label{fig:delta}
The quantity $\Delta$, eq.~(\ref{Delta}), is shown as a function of the
degree $n$ of the polynomial occurring in its definition. We compare 
the naive, pairing, bit reversal (br) and Montvay's
(M) ordering schemes. 
}}
\end{center}
\end{figure}

Since the Clenshaw recurrence is believed to be the numerically most stable
method to evaluate linear combinations of Chebyshev polynomials, 
the values of $\Delta$ can be interpreted as a measure for the effects 
of rounding errors in the evaluation of $\Phi_{\rm order}$. 
The results for $\Delta$ as a function of $n$ are shown in Fig.~\ref{fig:delta}. 
Using the naive ordering scheme, we could not run the cases of $n=64$
and $n=100$ 
because of 
numerical overflows. When adopting the pairing scheme, 
$\Delta$ takes large values for $n=64$ and $n=100$.
The bit-reversal scheme gives small but non-negligible values of
$\Delta$ for the considered values of $n$. Finally, Montvay's scheme
yields $\Delta $ of order $10^{-6}$ for all values of $n$. 
We conclude that, 
somewhat surprisingly,
it is possible to find ordering schemes through which 
the construction
of the polynomial $\hat{Q}^2P_{n,\epsilon}(\hat{Q}^2)$ can be done
with a precision that is comparable with the one expected when
performing $O(n)$ multiplications in 32-bit arithmetics. 
As already observed,  
one may evaluate the corresponding quantity $\Delta$ also for the product
representation eq.(\ref{pol_prod_q_start}) \cite{PHMC_hitech}. 
In this case again, small values of $\Delta \approx O(10^{-6})$ are found  
when using, however, {\em either} Montvay's or bit-reversal ordering schemes. 
This indicates that the improvement achieved
through these ordering schemes may also depend 
to some extent on the chosen product representation.

\section{32-bit versus 64-bit precision} 

In order to
obtain $\Delta$ in eq.(\ref{Delta}), the 
vector $\Phi_{\rm Clenshaw}$ has been computed
using solely 32-bit precision.
It might be asked therefore, whether 
$\Delta$ can really be considered as a measure of 
the size of the rounding errors occurring 
in the construction of $\Phi_{\rm order}$.
To the best of our knowledge, the theorem stated in \cite{foxy},
providing an upper bound on the rounding errors that occur
in the use of the Clenshaw recurrence, is not straightforwardly
generalizable to the case of a polynomial in a matrix.
Hence it can not be guaranteed that, in the case of a polynomial
in a matrix, the Clenshaw recursion 
is providing us 
with a reference method to quantify the rounding errors on $\Phi_{\rm order}$.
In order to  clarify this point, we decided to compare results
obtained from 32-bit precision with those obtained from 64-bit
precision programs. For this test we considered the SU(2) gauge group 
with the same
bare parameters as specified in
section 3.3, using three different lattice sizes. All the
results reported in this section were obtained on a single
thermalized gauge configuration and hence are given without a statistical
error. 
However, we checked explicitly in some cases that
using different gauge configurations yields only negligible changes in the
results discussed here. 
 
In the following, we denote by
\be \label{endvector}
 \chi_{\rm order} \equiv \hat{Q}\sqrt{c_{2n}} (\hat{Q}-r_{2n}) \cdot \dots
  \cdot \sqrt{c_1} (\hat{Q}-r_1) \hat{Q}R_G \nonumber \\
\ee
the vector obtained with
32-bit precision
and by $\tilde{\chi}_{\rm order}$
the corresponding vector obtained with 64-bit precision.
The subscript ``order'' labels again different ordering
schemes and $R_G$ is a random vector obtained from a Gaussian
distribution with unit variance. 
Analogously, we denote
by $\chi_{\rm Clenshaw}$ and $\tilde{\chi}_{\rm Clenshaw}$ the vectors obtained
using the Clenshaw recursion with the two precisions.

Then the norm of the difference between the vectors $\chi_{\rm order}$
and $\tilde{\chi}_{\rm order}$,
\be \label{eta_32_64}
\eta_{\rm order} =
\frac{1}{\sqrt{N}} \| \chi_{\rm order} - \tilde{\chi}_{\rm order}\|\; ,
\ee
with $N$ the number of degrees of freedom of the vector $\chi$,
can serve as an estimate of the rounding errors when 
$\chi_{\rm order}$ is evaluated.
Let us start by comparing the values of $\eta_{\rm order}$ for
the subpolynomial (sp), bit reversal (br) and Montvay's ordering schemes,
taking $n=64$ and $\epsilon=0.0015$:
\be \nn
\eta_{\rm sp} \simeq 3.7 \cdot 10^{-5} \quad , \quad
\eta_{\rm br} \simeq 4.3 \cdot 10^{-6} \quad , \quad
\eta_{\rm Montvay} \simeq 5.5 \cdot 10^{-6} \quad . \label{eta_su2}
\ee
The above values, in particular for the bit reversal and Montvay's ordering scheme,
are close to the precision level that is expected to be reached
optimally 
on 32-bit machines after
performing $2n+2 = 130$ multiplications.
 
Of particular interest are the values of $\eta_{\rm Clenshaw}$.
Since we expect that the vector $\chi$ is most precisely
evaluated when the Clenshaw recurrence relation is used with
64-bit precision, the values of $\eta_{\rm Clenshaw}$ 
would tell us how much the computation of $\chi_{\rm Clenshaw}$ is 
affected by rounding errors. 
The results are shown in table \ref{tab:clen_test}. 
We compare also $\chi_{\rm order}$ with
$\tilde{\chi}_{\rm Clenshaw}$, using the bit-reversal scheme
as an example for an ordering scheme and evaluate 
\be \label{etahat_32_64}
\hat{\eta}_{\rm br} = \frac{1}{\sqrt{N}}
\| \chi_{\rm br} - \tilde{\chi}_{\rm Clenshaw}\| \; ,
\ee
again reporting the results in table \ref{tab:clen_test}. 
 
\begin{table*}[hbt]
\begin{center}
\caption{The quantities $\eta_{\rm Clenshaw}$, eq.(\ref{eta_32_64}), 
and $\hat{\eta}_{\rm br}$, eq.(\ref{etahat_32_64}), for various
parameters of the polynomial $P_{n,\epsilon}$.
The values in the table are computed using
one thermalized SU(2) gauge configuration
at $\beta=1.75$, $\kappa=0.165$ and $c_{\rm sw}=0$.
Different lattices sizes $L^3\cdot T$ are compared.
}
\vspace{2mm}
\label{tab:clen_test}
\begin{tabular}{llllll}
\hline
$L^3\cdot T$ & $n$ & $\epsilon$ & $\delta$ & $\eta_{\rm Clenshaw}$ &
$\hat{\eta}_{br}$  \\
\hline \hline
$8^3\cdot 16$ & $16$ & $ 0.0215$ & $ 0.013$ & $8.4\cdot 10^{-8}$ &
$1.6\cdot 10^{-6}$   \\
$8^3\cdot 16$ & $32$ & $ 0.0058$ & $ 0.013$ & $1.3\cdot 10^{-7}$ &
$2.1\cdot 10^{-6}$   \\
$8^3\cdot 16$ & $64$ & $ 0.0015$ & $ 0.013$ & $2.7\cdot 10^{-7}$ &
$4.3\cdot 10^{-6}$    \\
$8^3\cdot 16$ & $100$ & $ 0.0006$ & $ 0.014$ & $6.0\cdot 10^{-7}$ &
$9.3\cdot 10^{-6}$    \\
\hline
$4^3\cdot 8 $ & $64$ & $ 0.0015$ & $ 0.013$ & $2.7\cdot 10^{-7}$ &
$4.6\cdot 10^{-6}$    \\
$6^3\cdot 16$ & $64$ & $ 0.0015$ & $ 0.013$ & $2.7\cdot 10^{-7}$ &
$4.5\cdot 10^{-6}$    \\
\hline
$8^3\cdot 16$ & $64$ & $ 0.0005$ & $ 0.109$ & $2.1\cdot 10^{-6}$ &
$5.6\cdot 10^{-6}$     \\
$8^3\cdot 16$ & $64$ & $ 0.0001$ & $ 0.545$ & $9.8\cdot 10^{-6}$ &
$8.7\cdot 10^{-6}$     \\
\hline \hline
\end{tabular}
\end{center}
\end{table*}
 
For practical values of $\delta < 0.014$ the Clenshaw recursion
relation turns out to be at least one order of magnitude
more precise than the bit-reversal scheme.
In fact, the magnitude of the rounding error using
the Clenshaw recursion looks in some cases even better 
than what is naively expected from
using 32-bit precision. This effect is easily explained by the fact
that the rounding error is suppressed by a factor of ${\rm O}(\delta)$,
as discussed in the previous section.
In addition, one can observe that, as $\delta$ increases,
the magnitude of the rounding error, as measured by
$\eta_{\rm Clenshaw}$, also grows and eventually reaches
values of the same order as for the bit-reversal scheme when
$\delta$ becomes of $O(1)$.
We conclude that for $\delta < 0.1$ the Clenshaw recursion
relation allows for a very precise evaluation of $\hat{Q}^2
P_{n,\epsilon}(\hat{Q}^2)$,
even when only 32-bit arithmetic 
is employed. 
This allows to test the 
different ordering schemes and to obtain reliable 
estimates of the rounding errors associated to
these schemes by using solely 32-bit precision. 
 
Table \ref{tab:clen_test} also shows that the magnitude of
rounding errors, as estimated by $\hat{\eta}_{br}$ for the
bit-reversal scheme, increases moderately with the degree
$n$ of the polynomial (with $\epsilon$ tuned to keep $\delta$
constant), but do not significantly depend on the considered
lattice sizes. The problem of the lattice size dependence of
rounding errors has not been systematically investigated.
However, on lattices
not larger than $8^3\cdot 16$, we could not observe any significant
dependence of the rounding
errors, measured through $\eta$ and $\hat{\eta}$, on the lattice size.                            
This might
be explained by the fact that $\hat{Q}$ has a band structure with
only the diagonal and some off-diagonal matrix elements not vanishing. 
For this argument to be valid, however, 
the precision losses, characterized by $\eta$, occurring in a single multiplication
by $\hat{Q}$, should be small enough for the propagation of their effect
through subsequent multiplications to be negligible. 
In practice we have found that when $\eta$ reaches a level 
of $\eta \approx O(10^{-4})$, 
a significant growth of the size of rounding errors, estimated by
$\eta$ itself, can be observed as the lattice volume increases. 
When such a behaviour sets in, 
it is certainly advisable to switch from
32-bit to 64-bit arithmetics.

\section{Conclusions}

In a class of fermion simulation algorithms, relying on the
multiboson technique, polynomials in a matrix have to be numerically
evaluated.
In a number of cases the polynomials are needed in their product representation,
in order to achieve an efficient performance of the algorithms.
Moreover, the simulations are often done on machines having only 32-bit 
precision. However, it is well known that, in evaluating a polynomial
using its product representation,
great care has to be taken, since rounding errors can easily lead
to significant precision losses and even numerical instabilities.
 
In this paper we investigated the 
effects of various ordering schemes for the monomial factors
or, equivalently, the complex roots 
in the numerical construction of a polynomial according to 
a product representation. We found that different ordering schemes
can lead to rounding-error effects ranging from numerical overflow
to retaining a precision 
comparable to the one that can be provided 
by the use of 
numerically stable
recurrence relations.
 
In the case of a polynomial of a single, real variable $s$,
approximating the function $1/s$, we introduced                         
estimators for the rounding errors, which give 
qualitative information about the level of precision losses 
occurring when 
a polynomial in a matrix is considered. 
These estimators indicate that the bit-reversal scheme
and a scheme suggested by Montvay can keep rounding-error effects to a low
level, for polynomials in a matrix of order up to $n \approx 200$.

Considering cases relevant for numerical simulations of lattice QCD, we studied
the magnitude of the precision losses affecting the evaluation of
polynomials in a particular matrix, when
adopting different ordering schemes for the monomials of the product
representation.
We found that Montvay's ordering scheme seems to be particularly
suited for this problem: by adopting this scheme, the rounding errors
could be kept at a level that is comparable to the one that is reached
when using the stable Clenshaw recurrence relation.
 
As the most important outcome of our investigation, we conclude that 
there exist 
orderings of the roots that allow a numerically precise evaluation
of a polynomial in a matrix, even up to degrees $n$ of the polynomial of ${\rm O}(100)$.
Although in this work only a particular matrix,
relevant for simulations of lattice QCD, has been considered,
we think that the main features of our results
should hold true for several different kinds of matrices and 
that, also, product representations can be used
in more general situations 
while keeping rounding-error effects
at a tolerable level. 
 
\vspace{0.5cm}
{\large\bf Acknowledgements}

This work is part of the ALPHA collaboration research programme. 
We are most grateful to S. Sint and U. Wolff for a critical reading of
the manuscript. 
We thank
DESY for allocating computer time to this project. R.F. thanks the 
Alexander von Humboldt Foundation for the financial support to his
research stay at DESY--Hamburg, where part of this work was done.

\clearpage

\begin{appendix}

\section{Definition of the matrix $\hat{Q}$} 
\label{gammas}

An application, where a product representation of 
polynomials in a matrix is used,
is the Monte Carlo simulation of fermion systems by the
multiboson technique and related algorithms. For completeness, 
we will give
in this appendix an explicit definition of the matrix 
$\hat{Q}$ used for the tests presented in this paper.
 
Let us consider 
a Euclidean space-time lattice with
lattice spacing $a$ and 
size $L^3\cdot T$. With the lattice spacing set to unity from now on, the
points on the lattice have integer coordinates $(t,x_1,x_2,x_3)$.     
A gauge field $U_{\mu}(x)\in SU(N_c)$ is assigned to the link
pointing from the site $x$ to the site $(x+\mu)$, where
$\mu=0,1,2,3$ denotes the four forward directions in space-time.
In this paper we will use $N_c=2,3$. For the case of $N_c=2$, 
the SU(2) gauge group, the boundary conditions are chosen
to be periodic in all directions. For $N_c=3$, the SU(3) gauge group,
we adopt periodic boundary conditions in space and Schr\"odinger
functional boundary conditions in the time direction \cite{paper3}.
After integrating out the fermion fields, their effects appear
in the resulting effective theory through the determinant of a
matrix $A=A[U]$, depending on the gauge field configuration $U$.
In the case of the $SU(N_c)$ gauge theory with $n_f=2$ mass degenerate
quark species, which is considered here, we have $A=Q^2$, where the matrix
$Q$, characterizing the fermion lattice action, is taken as follows:
\begin{eqnarray} 
\label{qmatrix}
Q[U]_{xy} \!\!\!&=& \!\!\!\frac{c_0}{c_M}\gamma_5 \Big[
(1+\sum_{\mu\nu}
[{i \over 2}c_{\rm sw}\kappa\sigma_{\mu\nu}{\cal F}_{\mu\nu}(x)])\delta_{x,y}
 \nonumber \\
&-&\kappa\sum_{\mu} \{
   (1-\gamma_{\mu})U_{\mu}(x)\delta_{x+\mu,y} +
(1+\gamma_{\mu})U^{\dagger}_{\mu}(x-\mu)\delta_{x-\mu,y}\} \Big]  \;\;.
\end{eqnarray} 
Here $\kappa$ and $c_{\rm sw}$ are parameters that have to be chosen
according to the physical problem under consideration,  
$c_0=(1+8\kappa)^{-1}$, and 
$c_M$ is a constant serving to optimize 
simulation algorithms. 
Typically $\kappa \approx 1/8$ and both $c_{\rm sw}$ and $c_M$ are 
${\rm O}(1)$. 

In order to speed up the Monte Carlo simulation,
it is  not the original matrix $Q$ 
but an even-odd preconditioned matrix $\hat{Q}$ that is used. 
Let us rewrite the matrix $Q$ in eq.~(\ref{qmatrix}) as
\begin{equation}
Q \equiv \frac{c_0}{c_M} \gamma_5  \left( \begin{array}{cc}
                1+T_{ee} & M_{eo} \\
                M_{oe} & 1+T_{oo} \\
                \end{array} \right)\;\;,
\end{equation} 
where we introduce the matrix $T_{ee}$($T_{oo}$) 
acting on vectors defined on the
even (odd) sites:
\begin{equation} \label{eq:t}
(T)_{xa\alpha,yb\beta} =
\sum_{\mu\nu}[{i \over 2}c_{sw}\kappa\sigma^{\alpha\beta}_{\mu\nu}
{\cal F}^{ab}_{\mu\nu}(x) \delta_{xy}]\;\;.
\end{equation}
The off-diagonal parts $M_{eo}$ and $M_{oe}$
connect the even with odd and odd with even lattice
sites, respectively.        
Preconditioning is now realized 
by writing the determinant of $Q$,
apart from an irrelevant constant factor, as
\begin{eqnarray} \label{preq}
\det(Q)&\propto&\det(1+T_{ee})\det\hat{Q}
\nonumber \\
\hat{Q}&=&\frac{\hat{c}_0}{c_M} \gamma_5(1 + T_{oo} - M_{oe}(1+T_{ee})^{-1}M_{eo})\;\;.
\end{eqnarray} 
The constant factor
$\hat{c}_0$ is given by $\hat{c}_0=1/(1+64\kappa^2)$,
and the constant $c_M$ 
is chosen such that the eigenvalues of $\hat{Q}$ are
in the interval $[-1,1]$. 

The matrices $\sigma_{\mu\nu}$, $\mu,\nu=0,...,3$ in the above expressions, are
defined by the commutator of $\gamma$-matrices
\begin{equation} \label{sigmamunu}
\sigma_{\mu\nu}=\frac{i}{2}\left[\gamma_\mu,\gamma_\nu\right]\;.
\end{equation}
The $4\otimes 4$ $\gamma$-matrices are given by
\begin{equation} \label{gamma}
\gamma_\mu = \left( \begin{array}{cc} 
                 0 & e_\mu \\
                 e_\mu^\dagger & 0 
              \end{array} \right)\;,
\end{equation} 
with the $2\otimes 2$ matrices 
\begin{equation}
e_0 = -1\; \; e_j = i\sigma_j,\; j=1,2,3\; 
\end{equation} 
and the Pauli-matrices $\sigma_j$
\begin{equation}
\sigma_1 = \left( \begin{array}{cc} 0 & 1 \\ 1 & 0 \end{array} \right)\;\;
\sigma_2 = \left( \begin{array}{cc} 0 & -i \\ i & 0 \end{array} \right)\;\;
\sigma_3 = \left( \begin{array}{cc} 1 & 0 \\ 0 & -1 \end{array} \right)\;\; .
\end{equation}
The matrix $\gamma_5 = \gamma_0\gamma_1\gamma_2\gamma_3$
is thus diagonal
\begin{equation} \label{gamma5}
\gamma_5 = \left( \begin{array}{cc}
                 1 & 0 \\
                 0 & -1
              \end{array} \right)\;.
\end{equation}

${\cal F}_{\mu\nu}(x)$ is a tensor antisymmetric under the exchange
$\mu \leftrightarrow \nu$ and, for given values of $\mu ,\nu$,   
an anti-Hermitian $N_c \otimes N_c$ matrix depending on the gauge links
in the vicinity of the lattice site $x$:
\pagebreak
\begin{eqnarray} 
{\cal F}_{\mu\nu}(x) &=& {1 \over 8} \left[
 U_{\mu}(x)  U_{\nu}(x+\hat{\mu})
U^{\dagger}_{\mu}(x+\hat{\nu})  U^{\dagger}_{\nu}(x)
\right.
\nonumber \\
&+& U_{\nu}(x) U^{\dagger}_{\mu}(x+\hat{\nu}-\hat{\mu})
U^{\dagger}_{\nu}(x-\hat{\mu})  U_{\mu}(x-\hat{\mu})
\nonumber \\
&+& U^{\dagger}_{\mu}(x-\hat{\mu}) U^{\dagger}_{\nu}(x-\hat{\nu}-\hat{\mu})
U_{\mu}(x-\hat{\nu}-\hat{\mu})  U_{\nu}(x-\hat{\nu})
\nonumber \\
&+& U^{\dagger}_{\nu}(x-\hat{\nu}) U_{\mu}(x-\hat{\nu})
U_{\nu}(x-\hat{\nu}+\hat{\mu})  U^{\dagger}_{\mu}(x)
\nonumber \\
&-& \left. {\rm h.c.} \right]\;\;.
\end{eqnarray} 

\end{appendix} 


\clearpage

\begin{figure}[t]

\setlength{\unitlength}{0.1bp}
\special{!
/gnudict 40 dict def
gnudict begin
/Color false def
/Solid false def
/gnulinewidth 5.000 def
/vshift -33 def
/dl {10 mul} def
/hpt 31.5 def
/vpt 31.5 def
/M {moveto} bind def
/L {lineto} bind def
/R {rmoveto} bind def
/V {rlineto} bind def
/vpt2 vpt 2 mul def
/hpt2 hpt 2 mul def
/Lshow { currentpoint stroke M
  0 vshift R show } def
/Rshow { currentpoint stroke M
  dup stringwidth pop neg vshift R show } def
/Cshow { currentpoint stroke M
  dup stringwidth pop -2 div vshift R show } def
/DL { Color {setrgbcolor Solid {pop []} if 0 setdash }
 {pop pop pop Solid {pop []} if 0 setdash} ifelse } def
/BL { stroke gnulinewidth 2 mul setlinewidth } def
/AL { stroke gnulinewidth 2 div setlinewidth } def
/PL { stroke gnulinewidth setlinewidth } def
/LTb { BL [] 0 0 0 DL } def
/LTa { AL [1 dl 2 dl] 0 setdash 0 0 0 setrgbcolor } def
/LT0 { PL [] 0 1 0 DL } def
/LT1 { PL [4 dl 2 dl] 0 0 1 DL } def
/LT2 { PL [2 dl 3 dl] 1 0 0 DL } def
/LT3 { PL [1 dl 1.5 dl] 1 0 1 DL } def
/LT4 { PL [5 dl 2 dl 1 dl 2 dl] 0 1 1 DL } def
/LT5 { PL [4 dl 3 dl 1 dl 3 dl] 1 1 0 DL } def
/LT6 { PL [2 dl 2 dl 2 dl 4 dl] 0 0 0 DL } def
/LT7 { PL [2 dl 2 dl 2 dl 2 dl 2 dl 4 dl] 1 0.3 0 DL } def
/LT8 { PL [2 dl 2 dl 2 dl 2 dl 2 dl 2 dl 2 dl 4 dl] 0.5 0.5 0.5 DL } def
/P { stroke [] 0 setdash
  currentlinewidth 2 div sub M
  0 currentlinewidth V stroke } def
/D { stroke [] 0 setdash 2 copy vpt add M
  hpt neg vpt neg V hpt vpt neg V
  hpt vpt V hpt neg vpt V closepath stroke
  P } def
/A { stroke [] 0 setdash vpt sub M 0 vpt2 V
  currentpoint stroke M
  hpt neg vpt neg R hpt2 0 V stroke
  } def
/B { stroke [] 0 setdash 2 copy exch hpt sub exch vpt add M
  0 vpt2 neg V hpt2 0 V 0 vpt2 V
  hpt2 neg 0 V closepath stroke
  P } def
/C { stroke [] 0 setdash exch hpt sub exch vpt add M
  hpt2 vpt2 neg V currentpoint stroke M
  hpt2 neg 0 R hpt2 vpt2 V stroke } def
/T { stroke [] 0 setdash 2 copy vpt 1.12 mul add M
  hpt neg vpt -1.62 mul V
  hpt 2 mul 0 V
  hpt neg vpt 1.62 mul V closepath stroke
  P  } def
/S { 2 copy A C} def
end
}
\begin{picture}(3600,2160)(0,0)
\special{"
gnudict begin
gsave
50 50 translate
0.100 0.100 scale
0 setgray
/Helvetica findfont 100 scalefont setfont
newpath
-500.000000 -500.000000 translate
LTa
600 251 M
0 1858 V
LTb
600 251 M
63 0 V
2754 0 R
-63 0 V
600 363 M
31 0 V
2786 0 R
-31 0 V
600 511 M
31 0 V
2786 0 R
-31 0 V
600 587 M
31 0 V
2786 0 R
-31 0 V
600 623 M
63 0 V
2754 0 R
-63 0 V
600 734 M
31 0 V
2786 0 R
-31 0 V
600 882 M
31 0 V
2786 0 R
-31 0 V
600 958 M
31 0 V
2786 0 R
-31 0 V
600 994 M
63 0 V
2754 0 R
-63 0 V
600 1106 M
31 0 V
2786 0 R
-31 0 V
600 1254 M
31 0 V
2786 0 R
-31 0 V
600 1330 M
31 0 V
2786 0 R
-31 0 V
600 1366 M
63 0 V
2754 0 R
-63 0 V
600 1478 M
31 0 V
2786 0 R
-31 0 V
600 1626 M
31 0 V
2786 0 R
-31 0 V
600 1701 M
31 0 V
2786 0 R
-31 0 V
600 1737 M
63 0 V
2754 0 R
-63 0 V
600 1849 M
31 0 V
2786 0 R
-31 0 V
600 1997 M
31 0 V
2786 0 R
-31 0 V
600 2073 M
31 0 V
2786 0 R
-31 0 V
600 2109 M
63 0 V
2754 0 R
-63 0 V
600 251 M
0 63 V
0 1795 R
0 -63 V
882 251 M
0 63 V
0 1795 R
0 -63 V
1163 251 M
0 63 V
0 1795 R
0 -63 V
1445 251 M
0 63 V
0 1795 R
0 -63 V
1727 251 M
0 63 V
0 1795 R
0 -63 V
2009 251 M
0 63 V
0 1795 R
0 -63 V
2290 251 M
0 63 V
0 1795 R
0 -63 V
2572 251 M
0 63 V
0 1795 R
0 -63 V
2854 251 M
0 63 V
0 1795 R
0 -63 V
3135 251 M
0 63 V
0 1795 R
0 -63 V
3417 251 M
0 63 V
0 1795 R
0 -63 V
600 251 M
2817 0 V
0 1858 V
-2817 0 V
600 251 L
LT0
2974 1626 D
3417 453 D
3276 455 D
3135 459 D
2994 462 D
2854 466 D
2713 471 D
2572 476 D
2431 482 D
2290 489 D
2149 497 D
2009 505 D
1868 515 D
1727 526 D
1586 539 D
1445 553 D
1304 568 D
1163 585 D
1023 603 D
882 622 D
741 642 D
600 663 D
LT1
2974 1526 A
3417 1079 A
3276 1087 A
3135 1095 A
2994 1103 A
2854 1112 A
2713 1121 A
2572 1130 A
2431 1140 A
2290 1150 A
2149 1161 A
2009 1172 A
1868 1184 A
1727 1196 A
1586 1209 A
1445 1222 A
1304 1236 A
1163 1251 A
1023 1266 A
882 1282 A
741 1299 A
600 1316 A
LT2
2974 1426 B
3417 1781 B
3276 1791 B
3135 1802 B
2994 1812 B
2854 1822 B
2713 1833 B
2572 1844 B
2431 1856 B
2290 1867 B
2149 1879 B
2009 1892 B
1868 1905 B
1727 1918 B
1586 1932 B
1445 1947 B
1304 1962 B
1163 1978 B
1023 1994 B
882 2011 B
741 2029 B
600 2047 B
stroke
grestore
end
showpage
}
\put(2854,1426){\makebox(0,0)[r]{146}}
\put(2854,1526){\makebox(0,0)[r]{86}}
\put(2854,1626){\makebox(0,0)[r]{30}}
\put(2008,51){\makebox(0,0){$s_{\rm min}$ in units of $\epsilon$}}
\put(100,1180){%
\special{ps: gsave currentpoint currentpoint translate
270 rotate neg exch neg exch translate}%
\makebox(0,0)[b]{\shortstack{$R_{\rm max}$}}%
\special{ps: currentpoint grestore moveto}%
}
\put(3417,151){\makebox(0,0){2}}
\put(3135,151){\makebox(0,0){1.8}}
\put(2854,151){\makebox(0,0){1.6}}
\put(2572,151){\makebox(0,0){1.4}}
\put(2290,151){\makebox(0,0){1.2}}
\put(2009,151){\makebox(0,0){1}}
\put(1727,151){\makebox(0,0){0.8}}
\put(1445,151){\makebox(0,0){0.6}}
\put(1163,151){\makebox(0,0){0.4}}
\put(882,151){\makebox(0,0){0.2}}
\put(600,151){\makebox(0,0){0}}
\put(540,2109){\makebox(0,0)[r]{1e+07}}
\put(540,1737){\makebox(0,0)[r]{1e+06}}
\put(540,1366){\makebox(0,0)[r]{100000}}
\put(540,994){\makebox(0,0)[r]{10000}}
\put(540,623){\makebox(0,0)[r]{1000}}
\put(540,251){\makebox(0,0)[r]{100}}
\end{picture}

\caption{
\label{fig:scale}
$R_{\rm max}$, eq.~(\ref{rmax}), 
is shown as a function of $s_{\rm min}/\epsilon$, where $s_{\rm min}$ is the
lower end of the interval $[s_{\rm min},1]$ from which
the values of $s$ are taken to compute $R_{\rm max}$. 
We show data for three polynomials of different degree,   
$n=30$, $n=86$ and $n=146$
at fixed relative fit accuracy
$\delta=0.001$, using the bit-reversal scheme. 
Although different in magnitude, $R_{\rm max}$ shows also
for the other schemes a very similar flat behaviour
as a function of $s_{\rm min}/\epsilon$. 
}
\end{figure}
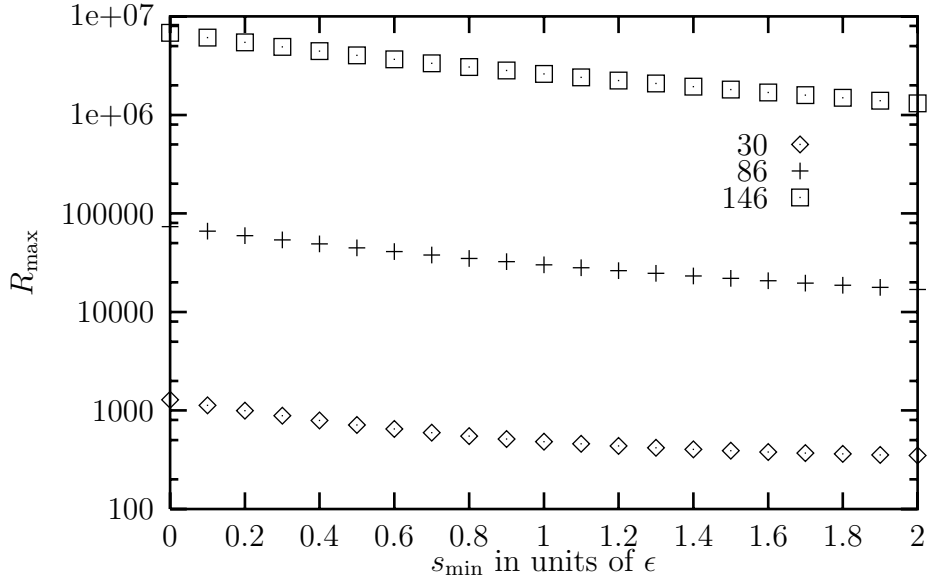

\begin{figure}[tbp]

\setlength{\unitlength}{0.1bp}
\special{!
/gnudict 40 dict def
gnudict begin
/Color false def
/Solid false def
/gnulinewidth 5.000 def
/vshift -33 def
/dl {10 mul} def
/hpt 31.5 def
/vpt 31.5 def
/M {moveto} bind def
/L {lineto} bind def
/R {rmoveto} bind def
/V {rlineto} bind def
/vpt2 vpt 2 mul def
/hpt2 hpt 2 mul def
/Lshow { currentpoint stroke M
  0 vshift R show } def
/Rshow { currentpoint stroke M
  dup stringwidth pop neg vshift R show } def
/Cshow { currentpoint stroke M
  dup stringwidth pop -2 div vshift R show } def
/DL { Color {setrgbcolor Solid {pop []} if 0 setdash }
 {pop pop pop Solid {pop []} if 0 setdash} ifelse } def
/BL { stroke gnulinewidth 2 mul setlinewidth } def
/AL { stroke gnulinewidth 2 div setlinewidth } def
/PL { stroke gnulinewidth setlinewidth } def
/LTb { BL [] 0 0 0 DL } def
/LTa { AL [1 dl 2 dl] 0 setdash 0 0 0 setrgbcolor } def
/LT0 { PL [] 0 1 0 DL } def
/LT1 { PL [4 dl 2 dl] 0 0 1 DL } def
/LT2 { PL [2 dl 3 dl] 1 0 0 DL } def
/LT3 { PL [1 dl 1.5 dl] 1 0 1 DL } def
/LT4 { PL [5 dl 2 dl 1 dl 2 dl] 0 1 1 DL } def
/LT5 { PL [4 dl 3 dl 1 dl 3 dl] 1 1 0 DL } def
/LT6 { PL [2 dl 2 dl 2 dl 4 dl] 0 0 0 DL } def
/LT7 { PL [2 dl 2 dl 2 dl 2 dl 2 dl 4 dl] 1 0.3 0 DL } def
/LT8 { PL [2 dl 2 dl 2 dl 2 dl 2 dl 2 dl 2 dl 4 dl] 0.5 0.5 0.5 DL } def
/P { stroke [] 0 setdash
  currentlinewidth 2 div sub M
  0 currentlinewidth V stroke } def
/D { stroke [] 0 setdash 2 copy vpt add M
  hpt neg vpt neg V hpt vpt neg V
  hpt vpt V hpt neg vpt V closepath stroke
  P } def
/A { stroke [] 0 setdash vpt sub M 0 vpt2 V
  currentpoint stroke M
  hpt neg vpt neg R hpt2 0 V stroke
  } def
/B { stroke [] 0 setdash 2 copy exch hpt sub exch vpt add M
  0 vpt2 neg V hpt2 0 V 0 vpt2 V
  hpt2 neg 0 V closepath stroke
  P } def
/C { stroke [] 0 setdash exch hpt sub exch vpt add M
  hpt2 vpt2 neg V currentpoint stroke M
  hpt2 neg 0 R hpt2 vpt2 V stroke } def
/T { stroke [] 0 setdash 2 copy vpt 1.12 mul add M
  hpt neg vpt -1.62 mul V
  hpt 2 mul 0 V
  hpt neg vpt 1.62 mul V closepath stroke
  P  } def
/S { 2 copy A C} def
end
}
\begin{picture}(3600,2160)(0,0)
\special{"
gnudict begin
gsave
50 50 translate
0.100 0.100 scale
0 setgray
/Helvetica findfont 100 scalefont setfont
newpath
-500.000000 -500.000000 translate
LTa
LTb
600 288 M
63 0 V
2754 0 R
-63 0 V
600 652 M
63 0 V
2754 0 R
-63 0 V
600 1016 M
63 0 V
2754 0 R
-63 0 V
600 1380 M
63 0 V
2754 0 R
-63 0 V
600 1745 M
63 0 V
2754 0 R
-63 0 V
600 2109 M
63 0 V
2754 0 R
-63 0 V
600 251 M
0 63 V
0 1795 R
0 -63 V
883 251 M
0 31 V
0 1827 R
0 -31 V
1048 251 M
0 31 V
0 1827 R
0 -31 V
1165 251 M
0 31 V
0 1827 R
0 -31 V
1256 251 M
0 31 V
0 1827 R
0 -31 V
1331 251 M
0 31 V
0 1827 R
0 -31 V
1394 251 M
0 31 V
0 1827 R
0 -31 V
1448 251 M
0 31 V
0 1827 R
0 -31 V
1496 251 M
0 31 V
0 1827 R
0 -31 V
1539 251 M
0 63 V
0 1795 R
0 -63 V
1822 251 M
0 31 V
0 1827 R
0 -31 V
1987 251 M
0 31 V
0 1827 R
0 -31 V
2104 251 M
0 31 V
0 1827 R
0 -31 V
2195 251 M
0 31 V
0 1827 R
0 -31 V
2270 251 M
0 31 V
0 1827 R
0 -31 V
2333 251 M
0 31 V
0 1827 R
0 -31 V
2387 251 M
0 31 V
0 1827 R
0 -31 V
2435 251 M
0 31 V
0 1827 R
0 -31 V
2478 251 M
0 63 V
0 1795 R
0 -63 V
2761 251 M
0 31 V
0 1827 R
0 -31 V
2926 251 M
0 31 V
0 1827 R
0 -31 V
3043 251 M
0 31 V
0 1827 R
0 -31 V
3134 251 M
0 31 V
0 1827 R
0 -31 V
3209 251 M
0 31 V
0 1827 R
0 -31 V
3272 251 M
0 31 V
0 1827 R
0 -31 V
3326 251 M
0 31 V
0 1827 R
0 -31 V
3374 251 M
0 31 V
0 1827 R
0 -31 V
3417 251 M
0 63 V
0 1795 R
0 -63 V
600 251 M
2817 0 V
0 1858 V
-2817 0 V
600 251 L
LT0
1659 1502 D
1165 449 D
1331 514 D
1331 530 D
1448 580 D
1539 623 D
1613 670 D
1676 687 D
1731 726 D
1822 775 D
1896 838 D
1959 892 D
2038 940 D
2104 978 D
2179 1025 D
2256 1116 D
2321 1129 D
2397 1216 D
2470 1260 D
2545 1305 D
2615 1382 D
2690 1436 D
2761 1513 D
2835 1615 D
2907 1675 D
2979 1761 D
3053 1868 D
3124 1991 D
LT1
1659 1402 A
1165 449 A
1331 462 A
1331 475 A
1448 580 A
1539 581 A
1613 594 A
1676 604 A
1731 726 A
1822 713 A
1896 743 A
1959 742 A
2038 930 A
2104 843 A
2179 886 A
2256 870 A
2321 1097 A
2397 1026 A
2470 1115 A
2545 983 A
2615 1185 A
2690 1096 A
2761 1168 A
2835 1052 A
2907 1257 A
2979 1162 A
3053 1238 A
3124 1183 A
stroke
grestore
end
showpage
}
\put(1539,1402){\makebox(0,0)[r]{Bit-reversal}}
\put(1539,1502){\makebox(0,0)[r]{Subpolynomial}}
\put(2008,51){\makebox(0,0){$n$}}
\put(100,1180){%
\special{ps: gsave currentpoint currentpoint translate
270 rotate neg exch neg exch translate}%
\makebox(0,0)[b]{\shortstack{$R_{\rm max}$}}%
\special{ps: currentpoint grestore moveto}%
}
\put(3417,151){\makebox(0,0){1000}}
\put(2478,151){\makebox(0,0){100}}
\put(1539,151){\makebox(0,0){10}}
\put(600,151){\makebox(0,0){1}}
\put(540,2109){\makebox(0,0)[r]{1e+15}}
\put(540,1745){\makebox(0,0)[r]{1e+12}}
\put(540,1380){\makebox(0,0)[r]{1e+09}}
\put(540,1016){\makebox(0,0)[r]{1e+06}}
\put(540,652){\makebox(0,0)[r]{1000}}
\put(540,288){\makebox(0,0)[r]{1}}
\end{picture}
\setlength{\unitlength}{0.1bp}
\special{!
/gnudict 40 dict def
gnudict begin
/Color false def
/Solid false def
/gnulinewidth 5.000 def
/vshift -33 def
/dl {10 mul} def
/hpt 31.5 def
/vpt 31.5 def
/M {moveto} bind def
/L {lineto} bind def
/R {rmoveto} bind def
/V {rlineto} bind def
/vpt2 vpt 2 mul def
/hpt2 hpt 2 mul def
/Lshow { currentpoint stroke M
  0 vshift R show } def
/Rshow { currentpoint stroke M
  dup stringwidth pop neg vshift R show } def
/Cshow { currentpoint stroke M
  dup stringwidth pop -2 div vshift R show } def
/DL { Color {setrgbcolor Solid {pop []} if 0 setdash }
 {pop pop pop Solid {pop []} if 0 setdash} ifelse } def
/BL { stroke gnulinewidth 2 mul setlinewidth } def
/AL { stroke gnulinewidth 2 div setlinewidth } def
/PL { stroke gnulinewidth setlinewidth } def
/LTb { BL [] 0 0 0 DL } def
/LTa { AL [1 dl 2 dl] 0 setdash 0 0 0 setrgbcolor } def
/LT0 { PL [] 0 1 0 DL } def
/LT1 { PL [4 dl 2 dl] 0 0 1 DL } def
/LT2 { PL [2 dl 3 dl] 1 0 0 DL } def
/LT3 { PL [1 dl 1.5 dl] 1 0 1 DL } def
/LT4 { PL [5 dl 2 dl 1 dl 2 dl] 0 1 1 DL } def
/LT5 { PL [4 dl 3 dl 1 dl 3 dl] 1 1 0 DL } def
/LT6 { PL [2 dl 2 dl 2 dl 4 dl] 0 0 0 DL } def
/LT7 { PL [2 dl 2 dl 2 dl 2 dl 2 dl 4 dl] 1 0.3 0 DL } def
/LT8 { PL [2 dl 2 dl 2 dl 2 dl 2 dl 2 dl 2 dl 4 dl] 0.5 0.5 0.5 DL } def
/P { stroke [] 0 setdash
  currentlinewidth 2 div sub M
  0 currentlinewidth V stroke } def
/D { stroke [] 0 setdash 2 copy vpt add M
  hpt neg vpt neg V hpt vpt neg V
  hpt vpt V hpt neg vpt V closepath stroke
  P } def
/A { stroke [] 0 setdash vpt sub M 0 vpt2 V
  currentpoint stroke M
  hpt neg vpt neg R hpt2 0 V stroke
  } def
/B { stroke [] 0 setdash 2 copy exch hpt sub exch vpt add M
  0 vpt2 neg V hpt2 0 V 0 vpt2 V
  hpt2 neg 0 V closepath stroke
  P } def
/C { stroke [] 0 setdash exch hpt sub exch vpt add M
  hpt2 vpt2 neg V currentpoint stroke M
  hpt2 neg 0 R hpt2 vpt2 V stroke } def
/T { stroke [] 0 setdash 2 copy vpt 1.12 mul add M
  hpt neg vpt -1.62 mul V
  hpt 2 mul 0 V
  hpt neg vpt 1.62 mul V closepath stroke
  P  } def
/S { 2 copy A C} def
end
}
\begin{picture}(3600,2160)(0,0)
\special{"
gnudict begin
gsave
50 50 translate
0.100 0.100 scale
0 setgray
/Helvetica findfont 100 scalefont setfont
newpath
-500.000000 -500.000000 translate
LTa
LTb
600 288 M
63 0 V
2754 0 R
-63 0 V
600 652 M
63 0 V
2754 0 R
-63 0 V
600 1016 M
63 0 V
2754 0 R
-63 0 V
600 1380 M
63 0 V
2754 0 R
-63 0 V
600 1745 M
63 0 V
2754 0 R
-63 0 V
600 251 M
0 63 V
0 1795 R
0 -63 V
883 251 M
0 31 V
0 1827 R
0 -31 V
1048 251 M
0 31 V
0 1827 R
0 -31 V
1165 251 M
0 31 V
0 1827 R
0 -31 V
1256 251 M
0 31 V
0 1827 R
0 -31 V
1331 251 M
0 31 V
0 1827 R
0 -31 V
1394 251 M
0 31 V
0 1827 R
0 -31 V
1448 251 M
0 31 V
0 1827 R
0 -31 V
1496 251 M
0 31 V
0 1827 R
0 -31 V
1539 251 M
0 63 V
0 1795 R
0 -63 V
1822 251 M
0 31 V
0 1827 R
0 -31 V
1987 251 M
0 31 V
0 1827 R
0 -31 V
2104 251 M
0 31 V
0 1827 R
0 -31 V
2195 251 M
0 31 V
0 1827 R
0 -31 V
2270 251 M
0 31 V
0 1827 R
0 -31 V
2333 251 M
0 31 V
0 1827 R
0 -31 V
2387 251 M
0 31 V
0 1827 R
0 -31 V
2435 251 M
0 31 V
0 1827 R
0 -31 V
2478 251 M
0 63 V
0 1795 R
0 -63 V
2761 251 M
0 31 V
0 1827 R
0 -31 V
2926 251 M
0 31 V
0 1827 R
0 -31 V
3043 251 M
0 31 V
0 1827 R
0 -31 V
3134 251 M
0 31 V
0 1827 R
0 -31 V
3209 251 M
0 31 V
0 1827 R
0 -31 V
3272 251 M
0 31 V
0 1827 R
0 -31 V
3326 251 M
0 31 V
0 1827 R
0 -31 V
3374 251 M
0 31 V
0 1827 R
0 -31 V
3417 251 M
0 63 V
0 1795 R
0 -63 V
600 251 M
2817 0 V
0 1858 V
-2817 0 V
600 251 L
LT0
1659 1502 D
1165 433 D
1331 469 D
1331 479 D
1448 524 D
1539 562 D
1613 596 D
1676 607 D
1731 656 D
1822 697 D
1896 733 D
1959 765 D
2038 806 D
2104 846 D
2179 889 D
2256 951 D
2321 978 D
2397 1034 D
2470 1064 D
2545 1100 D
2615 1152 D
2690 1176 D
2761 1213 D
2835 1268 D
2907 1302 D
2979 1345 D
3053 1403 D
3124 1457 D
LT1
1659 1402 A
1165 433 A
1331 460 A
1331 469 A
1448 524 A
1539 512 A
1613 539 A
1676 547 A
1731 656 A
1822 621 A
1896 673 A
1959 638 A
2038 839 A
2104 746 A
2179 810 A
2256 736 A
2321 973 A
2397 899 A
2470 1016 A
2545 793 A
2615 1045 A
2690 949 A
2761 1037 A
2835 898 A
2907 996 A
2979 970 A
3053 1076 A
3124 990 A
stroke
grestore
end
showpage
}
\put(1539,1402){\makebox(0,0)[r]{Bit-reversal}}
\put(1539,1502){\makebox(0,0)[r]{Subpolynomial}}
\put(2008,51){\makebox(0,0){$n$}}
\put(100,1180){%
\special{ps: gsave currentpoint currentpoint translate
270 rotate neg exch neg exch translate}%
\makebox(0,0)[b]{\shortstack{$M_{\rm max}$}}%
\special{ps: currentpoint grestore moveto}%
}
\put(3417,151){\makebox(0,0){1000}}
\put(2478,151){\makebox(0,0){100}}
\put(1539,151){\makebox(0,0){10}}
\put(600,151){\makebox(0,0){1}}
\put(540,1745){\makebox(0,0)[r]{1e+12}}
\put(540,1380){\makebox(0,0)[r]{1e+09}}
\put(540,1016){\makebox(0,0)[r]{1e+06}}
\put(540,652){\makebox(0,0)[r]{1000}}
\put(540,288){\makebox(0,0)[r]{1}}
\end{picture}
\caption{\label{fig:d0.1}
$R_{\rm max}$, eq.~(\ref{rmax}), 
and $M_{\rm max}$, eq.~(\ref{M}),
are shown as a function of the
degree of the polynomial
at fixed relative fit accuracy
$\delta=0.1$.
We compare subpolynomial and bit-reversal
ordering schemes.}
\end{figure}
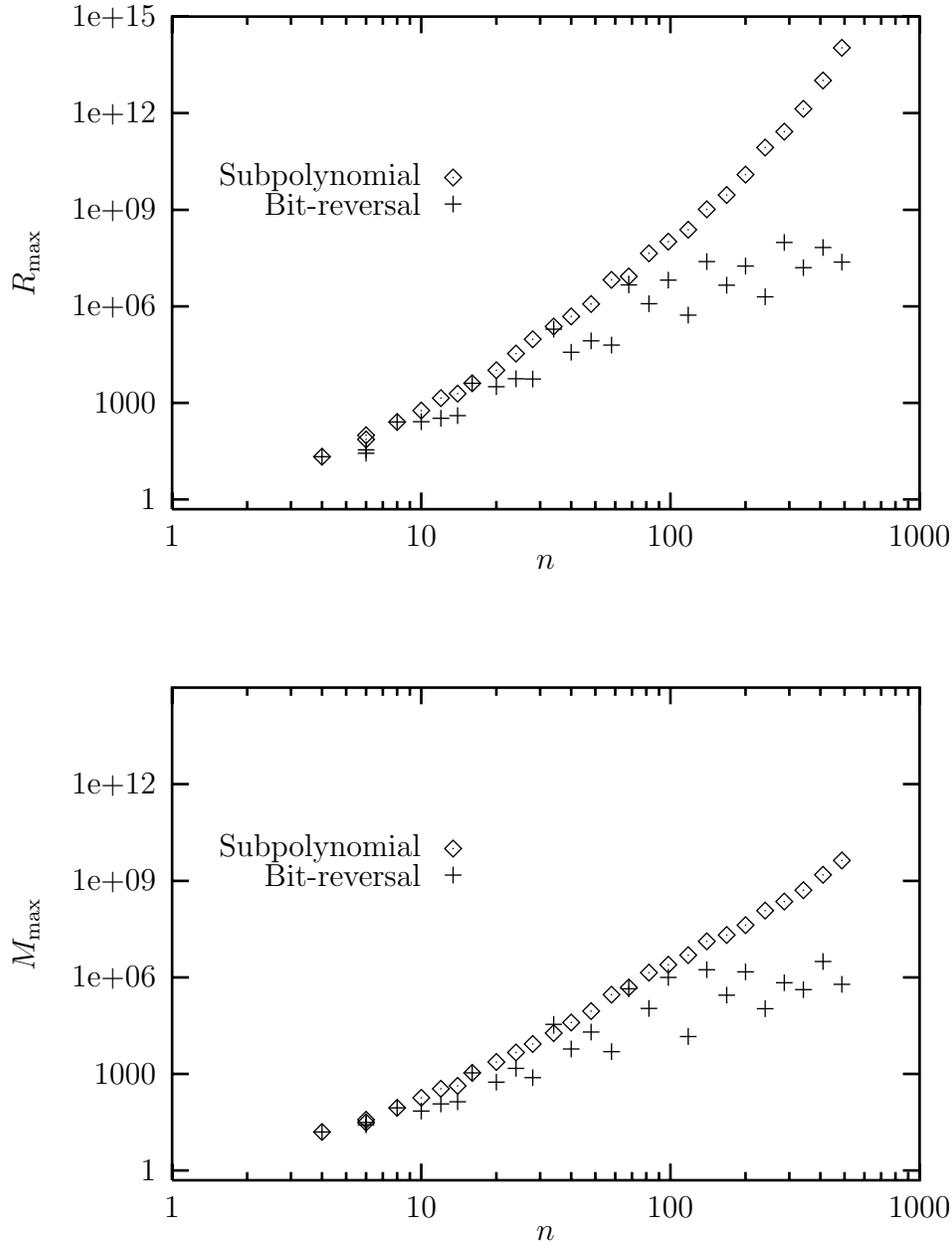

\end{document}